\documentclass[conference]{IEEEtran}
\IEEEoverridecommandlockouts
\usepackage{cite}
\usepackage{algorithmic}
\usepackage{graphicx}
\usepackage{textcomp}
\usepackage{xcolor}
\usepackage{enumitem,kantlipsum}
\usepackage{amssymb}
\usepackage{subfigure}
\usepackage{mathtools}
\usepackage{multirow}
\usepackage{pifont}
\usepackage{flushend}
\usepackage{xspace}

\def\BibTeX{{\rm B\kern-.05em{\sc i\kern-.025em b}\kern-.08em
    T\kern-.1667em\lower.7ex\hbox{E}\kern-.125emX}}

\begin{document}
\newcommand{\ours}{WiVelo\xspace}

\title{WiVelo: Fine-grained Walking Velocity Estimation for Wi-Fi Passive Tracking}

\author{
\IEEEauthorblockN{Chenning Li, Li Liu, Zhichao Cao, Mi Zhang}
\IEEEauthorblockA{Michigan State University \\
\{lichenni, liuli9, caozc, mizhang\}@msu.edu}
}


\maketitle

\begin{abstract}
Passive human tracking via Wi-Fi has been researched broadly in the past decade. 
Besides straight-forward anchor point localization, velocity is another vital sign adopted by the existing approaches to infer user trajectory.
However, state-of-the-art Wi-Fi velocity estimation relies on Doppler-Frequency-Shift (DFS) which suffers from the inevitable signal noise incurring unbounded velocity errors, further degrading the tracking accuracy. In this paper, we present \ours\footnote{Code\&datasets are available at 
\textit{https://github.com/liecn/WiVelo\_SECON22}} that explores new spatial-temporal signal correlation features observed from different antennas to achieve accurate velocity estimation.
First, we use subcarrier shift distribution (SSD) extracted from channel state information (CSI) to define two correlation features for direction and speed estimation, separately.
Then, we design a mesh model calculated by the antennas' locations to enable a fine-grained velocity estimation with bounded direction error.
Finally, with the continuously estimated velocity, we develop an end-to-end trajectory recovery algorithm to mitigate velocity outliers with the property of walking velocity continuity. 
We implement \ours on commodity Wi-Fi hardware and extensively evaluate its tracking accuracy in various environments. The experimental results show our median and 90\% tracking errors are 0.47~m and 1.06~m, which are half and a quarter of state-of-the-arts.
\end{abstract}

\begin{IEEEkeywords}
wireless sensing, Wi-Fi localization, channel state information, 
\end{IEEEkeywords}

\section{Introduction}
\label{sec-intro}

Human-centered sensing attracts massive interest with many new emerging applications (e.g., industry 4.0, smart health-care). Due to the importance of location-based services, human tracking~\cite{chen2019widesee, li2021deep} 
is broadly studied in the past decade. In indoor environments, human tracking becomes an essential task but is challenging due to a lack of infrastructure support. With the widely available and commercial Wi-Fi infrastructure, many approaches \cite{RIM, SpotFi, SAIL, Chronos} attach a Wi-Fi device actively on a human for device-based tracking. Although the active device can provide the centimeter-level tracking accuracy with a high signal resolution, it is infeasible in many scenarios due to the intrusive deployment and battery recharge.


Many \emph{passive tracking} approaches~\cite{IndoTrack, Widar, Music, CARM, Widar2.0, md-Track, WiTrack, LiFS, karanam2019tracking, xdTrack} eliminate the embedded Wi-Fi device by extracting different information from the signal bouncing off the target person to infer the trajectory. Power attenuation (PA), Angle of Arrival (AoA), Time of Flight (ToF), and Doppler Frequency Shift (DFS) provide independent ranging information. In comparison with others that directly link with a location's coordinates, DFS enables velocity estimation which is adopted by many approaches~\cite{Widar,Widar2.0,md-Track,IndoTrack}. 
However, DFS only recovers the velocity component along the direction determined by the relative position between users and transceivers, which usually makes the velocity errors unbounded in practice~\cite{Widar,Widar2.0}.

In this paper, we propose \ours that utilizes the spatial-temporal signal correlation among different antennas to achieve fine-grained velocity estimation, which further improves the accuracy of Wi-Fi passive tracking. Specifically, we divide the velocity estimation problem into two sub-problems. One is walking direction estimation, and the other is to calculate the speed along the direction. We solve these two sub-problems by using our defined spatial-temporal signal correlation, in which we use the channel state information (CSI) observed from different antennas to define a temporal signal correlation of an individual antenna and a spatial signal correlation between two antennas. 

\begin{figure}
\centering

    \includegraphics[width=0.35\textwidth]{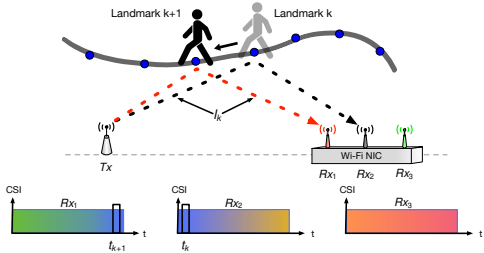}
    \vspace{-3mm}
	\caption{An illustration of the conceptual idea that explores spatial-temporal signal correlation to estimate walking velocity.}
	\label{fig-correlation-illustration}
	\vspace{-5mm}
\end{figure}

Figure~\ref{fig-correlation-illustration} illustrates the high-level idea of \ours including what our signal correlation refers to and how to connect the signal correlation with a user's walking velocity.
A series of landmarks (e.g., blue points) pinpoint the trajectory. A user is walking from landmark $k$ to $k+1$, and the corresponding arrival times are $t_k$ and $t_{k+1}$. 
Meanwhile, a Wi-Fi network interface card (NIC), consisting of 3 antennas (e.g., $Rx_1$, $Rx_2$, and $Rx_3$), continuously records CSI information (e.g., the color series at the bottom) during the user's movement. These antennas share an antenna $Tx$ as the signal transmitter. \\
\textbf{Walking Direction:} $Rx_2$ observes a ``blue" signal pattern in a short time window near $t_k$ (e.g., bottom-middle), referring to the temporal signal correlation that can be used to infer the walking direction. \\
\textbf{Walking Speed:} Later on, another antenna $Rx_1$ observes the same ``blue" signal pattern at $t_{k+1}$, referring to spatial signal correlation. According to the reflected path constraint, we use the coordinates of $Tx$, $Rx_1$, $Rx_2$ and the landmark $k$ to calculate the walking distance in the time period $t_{k+1} - t_{k}$. Thus, the walking speed can be calculated.

We use Subcarrier Shift Distribution (SSD), extracted from the CSI series, to calculate the temporal-spatial signal correlation. The sign of an SSD is sensitive to the length change of the signal reflection path ($\S$\ref{sec-preliminary}) determined by the walking direction. Thus, it allows us to define the temporal and spatial signal correlation for walking direction and speed estimation.
To mitigate the negative influence of multi-path noises on each SSD calculation, we propose three countermeasures.
First, we use the two pairs of Wi-Fi transceivers to develop a mesh model that can achieve fine-grained direction resolution and distance calculation ($\S$~\ref{sec-geometrical-model}).
Additionally, we utilize multiple subcarriers in Wi-Fi and multiple Wi-Fi packets received by an antenna in a short time to construct massive voters, who calculate the temporal signal correlation together for determining the walking direction ($\S$~\ref{sec-walk-direction}). 
Finally, given two SSDs calculated by two antennas, we use Earth Mover's Distance~\cite{EMD} to depict the distribution similarity as the spatial signal correlation for prohibiting the influence of various noises during speed estimation ($\S$~\ref{sec-anchor-point-hopping}).
With the estimated velocity, we develop a trajectory tracking system. We notice that the speed is continuous during walking. With the constraint, we jointly refine a series of velocity estimations during recovering the whole walking trajectory ($\S$~\ref{sec-trajectory-recovery}).

We implement \ours with three laptops equipped with Intel 5300 NIC~\cite{csitools}, delivering two Wi-Fi transceiver pairs. Extensive experiments are conducted to evaluate its performance for different users and trajectories in three environments. The results show that the median tracking error is 0.47~m and the 90-percentile tracking error is 1.06~m, which are 40.8\% and 26.5\% of state-of-the-arts. Our contributions are summarized as follows:
\begin{itemize}[leftmargin=*]
	\item We develop a fine-grained walking velocity estimation system with temporal-spatial signal correlation thus can achieve more accuracy for Wi-Fi passive tracking in complex scenarios.
	\item We design efficient direction and speed estimation algorithms to mitigate the SSD errors incurred by multi-path noises.
	\item We implement \ours with commodity Wi-Fi hardware and conduct extensive experiments in different scenarios. The results show \ours reduces the median and 90-percentile tracking errors by 59.2\% and 73.5\%.
\end{itemize}



\section{Preliminary}
\label{sec-preliminary}




\subsection{Channel State Information}
\label{subsec-CSI}

In \ours, the Wi-Fi signal transmitter broadcasts data packets continuously. 
Given various time (packet), frequency (subcarrier) and antenna (receiver) indicated by $t$, $f$ and $a$, raw CSI is defined as the sum of Channel Frequency Response (CFR) over multiple paths (e.g., Line-of-Sight (LoS), reflection, scattering)~\cite{Widar2.0, csi_survey} as follows:
\begin{equation}
\label{equ-csi}
\begin{split}
H(t,f,a)&=H_h(t,f,a)+\sum_{l=0}^L{H_l(t,f,a)} \\
&\triangleq{\alpha_h(t,f,a)e^{-j2\pi f\tau_h(t,f,a)}+N(t,f,a)} \\
H_l(t,f,a)&\triangleq \alpha_l(t,f,a)e^{-j2\pi f\tau_l(t,f,a)}, \forall l\in{[0,L]}
\end{split}
\end{equation}
where $H_h$ indicates the signal reflected by the human body. $\alpha_h$ and $\tau_h$ are the complex attenuation factor and propagation delay of $H_h$ respectively. $H_l$ is a signal from another path (e.g., LoS propagation, furniture reflection/scattering), which is a noise for $H_h$. We assume the LoS signal is $H_0$. For each noise path $H_l$, $\alpha_l$ and $\tau_h$ are the corresponding complex attenuation factor and propagation delay. Besides $H_0$, there are total $L$ noise paths, which can be focused on the 2 main paths (e.g., LoS and the reflected path from the targets) in the following process. We further combine these noises together, which is indicated as $N$. 

\subsection{Subcarrier Shift Distribution}
\label{subsec-subcarrier-shift-distribution}
State-of-the-art Wi-Fi physical layer usually adopts OFDM (Orthogonal Frequency Division Modulation) to transmit signals simultaneously through dozens of subcarriers operating at different orthogonal frequencies. 
According to Equation~\ref{equ-csi}, given a subcarrier (frequency $f$) and an antenna $a$, at time $t$, the power of the CFR $H(t)$ can be calculated as follows:
\begin{align}
\label{equ-energy} 
|H(t)|^2 =& |\alpha_h(t)|^2 + |N(t)|^2 + 2\sum^L_0|\alpha_h(t)\alpha_l(t)|\cos \rho_l(t)\nonumber\\
\rho_l(t) =& 2\pi f(\tau_h(t)-\tau_l(t))+\phi_l, \forall l \in [0,L]
\end{align}
where $\phi_l$ is a constant phase offset induced by electric polarization~\cite{InitPhase, WiDir}. $\rho_l$ depicts the total phase offset between $H_h$ and $H_l$. Moreover, by filtering out the minor power contributed by the non-LoS noises with Savitzky-Golay finite impulse response smoothing Filter (SGF)~\cite{SGF,WiDir}, the CFR power $|H(t)|^2$ is approximately determined by LoS signal and human reflected signal as follows:
\begin{equation}
\label{equ-power-filtering}
\begin{split}
|H(t)|^2 \approx & |\alpha_0(t)|^2 + 2|\alpha_0(t) \alpha_h(t)|\cos{(\rho_0(t))} \\
\rho_0(t) =& 2\pi f(\tau_h(t)-\tau_0(t))+\phi_0
\end{split}
\end{equation}
where $\tau_0$ is deterministic due to the static LoS path. When the human body is moving, $\tau_h$ maybe continuously change so that the CFR power must have a fluctuated pattern in the time domain. 

\begin{figure}
\centering
    \includegraphics[width=0.3\textwidth]{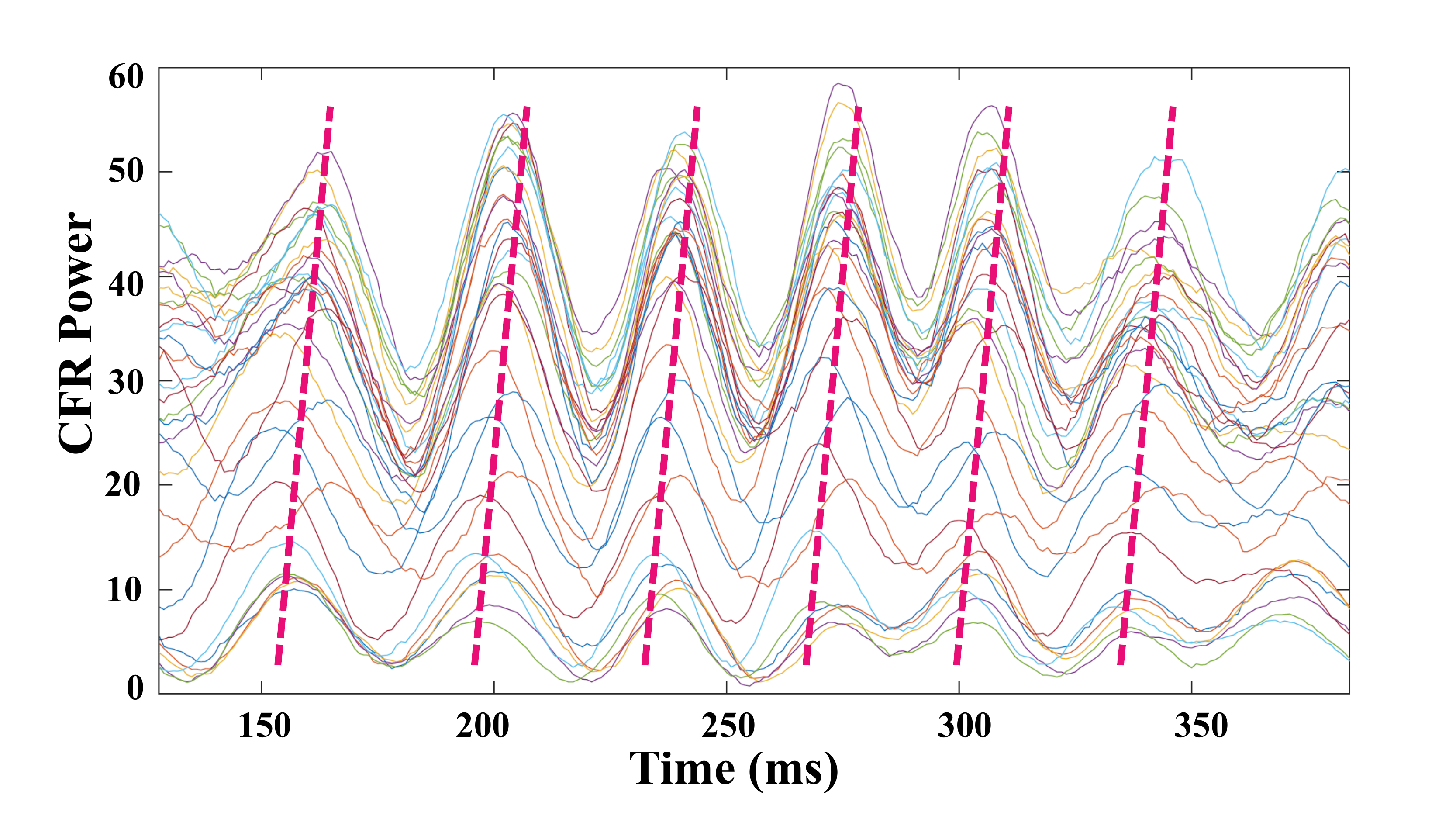}
    \vspace{-2mm}
    \caption{Fluctuation patterns of all subcarriers' CFR power $|H(t)|^2$ during a walk. Each curve represents the pattern of a subcarrier.}
    \label{fig-ssd}
\vspace{-5mm}
\end{figure}

We use a pair of transceivers equipped with Intel 5300 NIC to collect CSI measurements in a walk. Figure~\ref{fig-ssd} shows the fluctuation patterns of different subcarriers' CFR power~\cite{LiFS} 
in the walk. As expected, we clearly see a sinusoid pattern for all subcarriers. Moreover, the dashed lines show a steady time offset of the maximum CFR power across all subcarriers. 
The time offset between two subcarriers is determined by the phase offset and walk direction.
Specifically, the length of LoS path and human reflected path is $d_0$ and $d_h$, respectively. We assume $|\alpha_0|^2$ is constant in short-term. For antenna $a$, given two different subcarriers $f_1$ and $f_2$, at time $t$, the difference of the phase offset $\rho_0(f_1,t)$ and $\rho_0(f_2,t)$ is:
\begin{equation}
\label{equ-phase-offset} 
\begin{split}
\Delta \rho(f_1,f_2|t) &=\frac{2\pi (d_h(t)-d_0(t))}{c/(f_{1}-f_{2})}
\end{split}
\end{equation}
Where $c$ is light speed. If the $f_1-f_2$ is fixed, the phase offset only relates to the length variation of the human reflected path. Given two opposite walking directions~\cite{WiDir}, one increases the length of the human reflected path, but the other decreases, the phase offset can be compensated with an opposite time offset along the time axis.
We denote the series of observed time offsets between multiple pairs of subcarriers as {\itshape Subcarrier Shift Distribution} (SSD) vector at time $t$. SSD vector is resilient to multiple types of random phase offsets~\cite{csi_survey} induced by the unsynchronized transceivers if it is computed for the individual antenna or multiple antennas in the same NIC.

\section{Overview}
\label{sec-overview}


\begin{figure}
\centering
\includegraphics[width=0.35\textwidth]{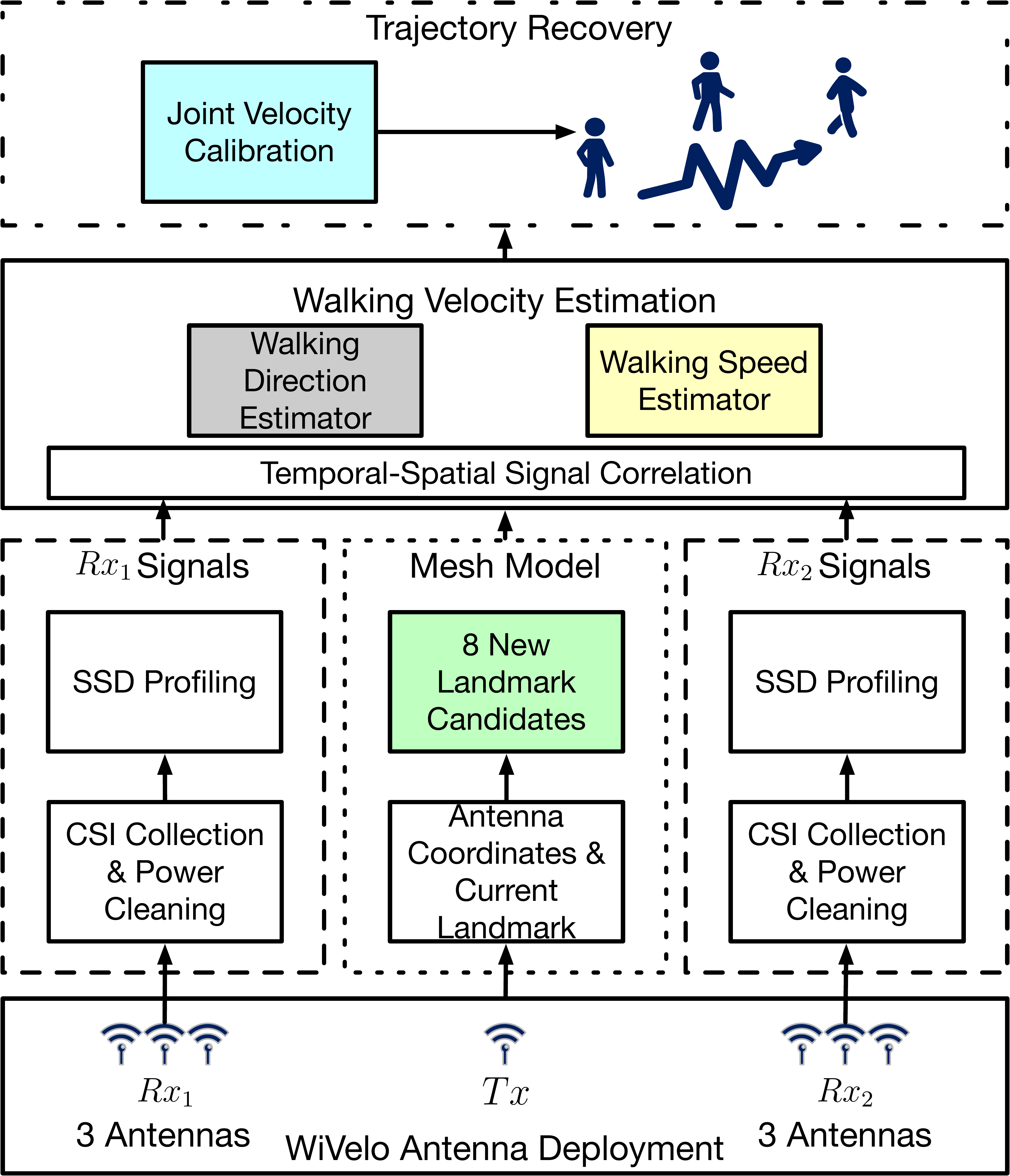}
\vspace{-2mm}
\caption{System overview of \ours with four parts in different colors.}
\label{fig-overveiw}
\vspace{-5mm}
\end{figure}

\ours contains four parts (e.g. mesh model generator, walking direction estimator, walking speed estimator, and trajectory recovery) denoted with various colors in Figure~\ref{fig-overveiw}.

\begin{enumerate}[wide, labelwidth=!, labelindent=0pt]
\vspace{1mm}
\item \textbf{Mesh Model Generator ($\S$\ref{sec-geometrical-model}).} The coordinates of all transceiver antennas are pre-configured. Given the coordinate of the current position, we calculate the reference length of the two reflection paths between the Tx-antenna and two references Rx-antennas, respectively. Then, according to the constraints of temporal-spatial signal correlation, we combine the coordinates of some auxiliary Rx-antennas and the reference lengths to derive the mesh model, which consists of 8 possible landmarks towards 8 different walking directions. 

\vspace{1mm}
\item \textbf{Walking Direction Estimator ($\S$\ref{sec-walk-direction}).} In a short period (e.g. 128 ms), we collect CSI measurements from all Rx-antennas and remove the noise from CFR power with SGF. Then, we profile SSDs from a series of CFR power for each Rx-antenna at a different time (packet). According to the observed temporal signal correlation, we estimate its walking direction.

\vspace{1mm}
\item \textbf{Walking Speed Estimator ($\S$\ref{sec-anchor-point-hopping}).}
In the mesh model, the walking direction is determined by the walking direction status of the two reference antennas. Then, given the walking direction, we determine the landmark. Additionally, we define the spatial signal correlation between two Rx-antennas to search the walking time interval. Finally, the walking speed can be calculated with the walking distance and the time interval. 

\vspace{1mm}
\item \textbf{Trajectory Recovery ($\S$\ref{sec-trajectory-recovery}).}
Upon receiving a sequence of walking velocities, which are treated as a constant in a short time slice (e.g., 128 ms), we jointly calibrate the velocity estimation, then connect all the displacements in all short periods to recover the trajectory.
\end{enumerate}

\section{Fine-grained Mesh Model}
\label{sec-geometrical-model}

\begin{figure}
\centering
\includegraphics[width=0.3\textwidth]{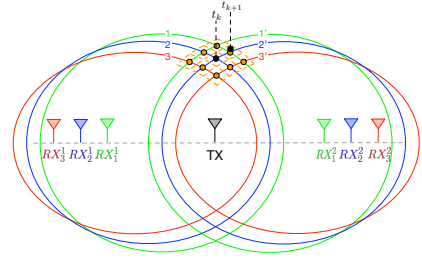}
\vspace{-2mm}
\caption{The illustration of the mesh model. We generate 8 new position candidates as the intersections of two groups of correlated ellipses}
\label{fig-geometric-model}
\vspace{-5mm}
\end{figure}


In \ours, SSD has two properties that make it possible to estimate walking velocity.
\begin{itemize}
    \item \textbf{Walking Direction:} SSD can identify whether the reflection path length is increasing or decreasing. Given the positions of the transceiver antennas, the trend can be further used to estimate the walking direction.
    \item \textbf{Walking Speed:} Two Rx-antennas may observe similar SSD at different times. During the time interval, the walking distance can be inferred by the geometrical constraint that their reflection paths have the same length.
\end{itemize}


As shown in Figure~\ref{fig-geometric-model}, \ours consists of two pairs of Wi-Fi transceivers. Two Wi-Fi NICs are the receivers (e.g., $Rx^1$, $Rx^2$), each has 3 Rx-antennas. They share the same Tx-antenna (e.g., $Tx$) equipped on another Wi-Fi NIC. 
The 6 Rx-antennas are deployed at both sides of the Tx-antenna symmetrically. 
These Rx-antennas provide two kinds of functionalities as follows:

    \noindent
    \textbf{Reference Antenna:}
    For each receiver, its middle Rx-antenna (e.g., $Rx^1_2$, $Rx^2_2$) is called \emph{reference antenna}. 
    From the view of a reference antenna, the walking direction has three possibilities.
    If its SSD is zero, the walking direction is along an ellipse (e.g., 2 for $Rx^1_2$, 2' for $Rx^2_2$), called \emph{reference ellipse}, determined by the positions of the reference antenna and the Tx-antenna, and its reflection path length, called \emph{reference length}.
    Otherwise, its SSD sign can determine the walking direction is towards the inside or outside of the ellipse with the first SSD property.
    We combine the two reference antennas' observations to obtain 8 possible walking directions so that achieve a 22.5$^\circ$ direction resolution.
    
    \noindent
    \textbf{Auxiliary Antenna:}
    For each receiver, besides the reference antenna, the other two Rx-antennas (e.g., $Rx^1_1$/$Rx^1_3$, $Rx^2_1$/$Rx^2_3$) are called \emph{auxiliary antennas}. 
    Given the reference length, and the positions of the auxiliary antennas and the Tx-antenna, we calculate another four ellipses (e.g., 1/3 for $Rx^1_1$/$Rx^1_3$, 1'/3' for $Rx^2_1$/$Rx^2_3$), called \emph{auxiliary ellipses}. 
    Two reference ellipses and four auxiliary ellipses intersect at 9 positions forming a \emph{mesh model}, which consist of the current position (e.g., the intersection between 2 and 2') and 8 possible arrival positions that correspond to the 8 possible walking directions observed by the reference antennas.
    Then, according to the estimated walking direction, we determine which arrival position (called \emph{landmark}) is on the direction and select the corresponding auxiliary antennas to calculate the walking speed with the second SSD property.

For example, 
at time $t_k$, a user's position is the inner black dot in Figure~\ref{fig-geometric-model}.
The SSD observed by a reference antenna $Rx^2_2$ indicates the user is staying on the reference ellipse 2'. And the SSD observed by the other reference antenna $Rx^1_2$ indicates the user is moving to the outside of the reference ellipse 2. We know that the walking direction is towards the landmark, which is the intersection between the auxiliary ellipse 1 and the reference ellipse 2'.
Then, we search the SSD observed by the auxiliary antenna $R^1_1$ to determine a time $t_{k+1}$ when it is the most similar with the SSD observed by $Rx^1_2$ at time $t_k$. Additionally, $t_{k+1}$ is the time when the user arrives at the landmark.
Since the coordinates of the 9 intersections are known, the walking velocity can be calculated by obtaining the time interval $t_{k+1}-t_k$.
Actually, the user may arrive at the other black dot near the landmark at time $t_{k+1}$. However, the maximum walking direction error is approximate 22.5$^\circ$, and the relative error of walking speed is less than 8.2\% (i.e., $1/\cos{22.5^\circ}-1$).
In addition, given the coordinates of current position, Tx-antenna, reference antennas, and auxiliary antennas, we can generate two groups of ellipses to calculate the coordinates of the intersections in our mesh model.

ilamp\section{Walk Direction Estimator}
\label{sec-walk-direction}

In this section, we step forward to resolve the walking direction estimation problem with a short-term series of CSI measurements obtained from receiver antennas. Given the time period $T$ of the CSI sequence and the arrival time $t_k$ of the start landmark. We empirically set $T$ to ensure it is a little larger than all possible $t_{k+1}-t_k$ according to the maximum distance between two adjacent landmarks in a monitoring area and the normal human walking speed.
%
%
Given the time period $T$, after collecting the raw CSI sequences from all receiver antennas, we filter out the CFR power noise with SGF. Then, we select $n$ different pairs of subcarriers to generate the SSD vector $\vec{S}(t,a) = [SSD_i|t,a]^T,i\in[1,n]$ for a receiver antenna $a$, at time (packet) $t$ ($t \in [0, T]$). 

To enhance the reliability of the walking direction estimation, we compute it in two steps. First, we define a temporal signal correlation pattern, namely {\itshape sign distance}, between two SSD vectors to distinguish whether the user moves away from the current ellipse. Given a time (packet) interval $\Delta t$, the sign distance between two SSD vectors $\vec{S}(t,a)$ and $\vec{S}(t+\Delta t,a)$  can be formulated as follows:
\begin{equation}
\label{equ-sign-distance} 
\begin{split}
\kappa(t,a)=\frac{Sign(\vec{S}(t,a)^T)\cdot Sign(\vec{S}(t+\Delta t,a))}{|Sign(\vec{S}(t,a))||Sign(\vec{S}(t+\Delta t,a))|}
\end{split}
\end{equation}
Where $Sign(\vec{S}(t,a))$ indicates the sign vector of each element in SSD vector $\vec{S}(t,a)$. Intuitively, we can define a threshold. If $\kappa(t, a)$ is smaller than the threshold, the user is still on the current ellipse. Otherwise, the user moves away from the current ellipse.

Our observation is that due to the SSD noise, even the user does not move away from the current ellipse, the sign of different SSD elements randomly fluctuated between negative and positive, which will significantly mislead the walking direction estimation if we only count the majority sign of an SSD vector. By utilizing the randomness, which means the elements of $Sign(\vec{S}(t, a))$ may have different signs with the corresponding element in $Sign(\vec{S}(t+\Delta t, a))$, adding the positive and negative together makes $\kappa(t, a)$ tend to equal zero when the user does not move away from the current ellipse. In the opposite, the elements of $Sign(\vec{S}(t,a))$ have the signs with the corresponding element in $Sign(\vec{S}(t+\Delta t,a))$. As a result, positive values continuously accumulate to $\kappa(t,a)$.

In the second step, if the user does not move away from the current ellipse, we skip this step. Otherwise, the majority signs of $Sign(\vec{S}(t,a))$ and $Sign(\vec{S}(t+\Delta t,a))$ are utilized to determine the walking direction (e.g., walk towards or move away from the antenna $a$).


In the time period $T$, we further use the redundant signals to construct many pairs of SSD vectors as \emph{massive virtual voters} to suppress the noise of individual pairs of SSD vectors, which leads negative influence on the judgment of whether the user moves away from the current ellipse. First, we split the whole time period $T$ into $N$ time windows, which have the same time span $T_w$. The $N$ time windows are uniformly distributed in the time period $T$ and partially overlapped. In each time window, antenna $a$ totally receives $M$ packets, then we have total $M$ different SSD vectors indicated as $\vec{S}(i, a), i\in[1,M]$. Then, give a packet (time) delay $\delta$, we have total $M-\delta$ pairs of SSD vectors as $(\vec{S}(i,a),\vec{S}(i+\delta,a), i\in[1,M-\delta]$. The $\delta$ should be large enough to avoid the interference of a similar noise pattern between time-close SSD vectors. For each pair, we can obtain a sign distance $\kappa(i,a)$ according to Equation~\ref{equ-sign-distance}, where $t$ and $\Delta t$ correspond to $i$ and $\delta$. Overall, we have total $N \times (M-\delta)$ virtual voters to determine the two-step walk direction estimation with majority decision.

\section{Walking Speed Estimator}
\label{sec-anchor-point-hopping}

\begin{figure*}[htp]
\centering
\subfigure{\includegraphics[width=0.195\textwidth]{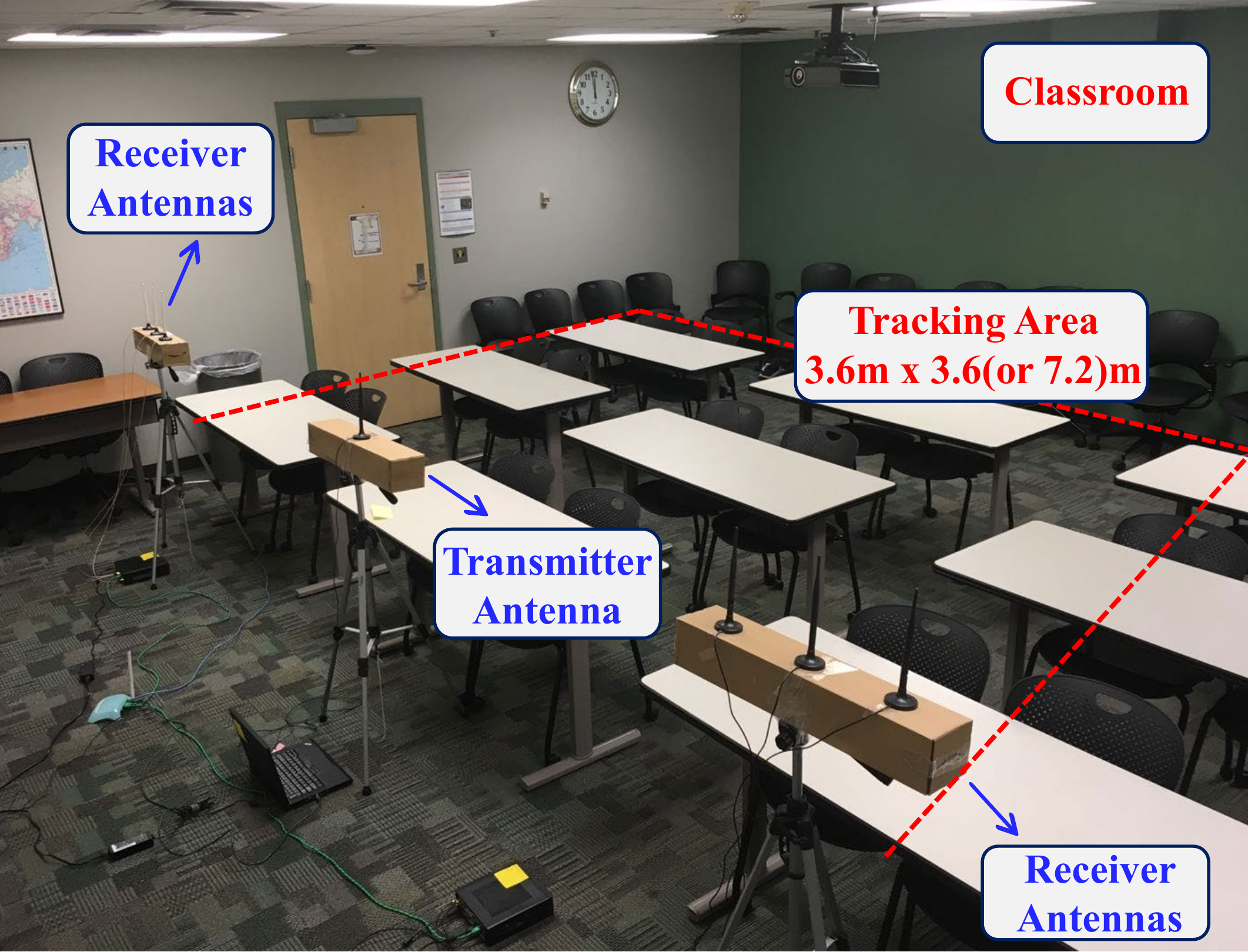}}
\subfigure{\includegraphics[width=0.3\textwidth]{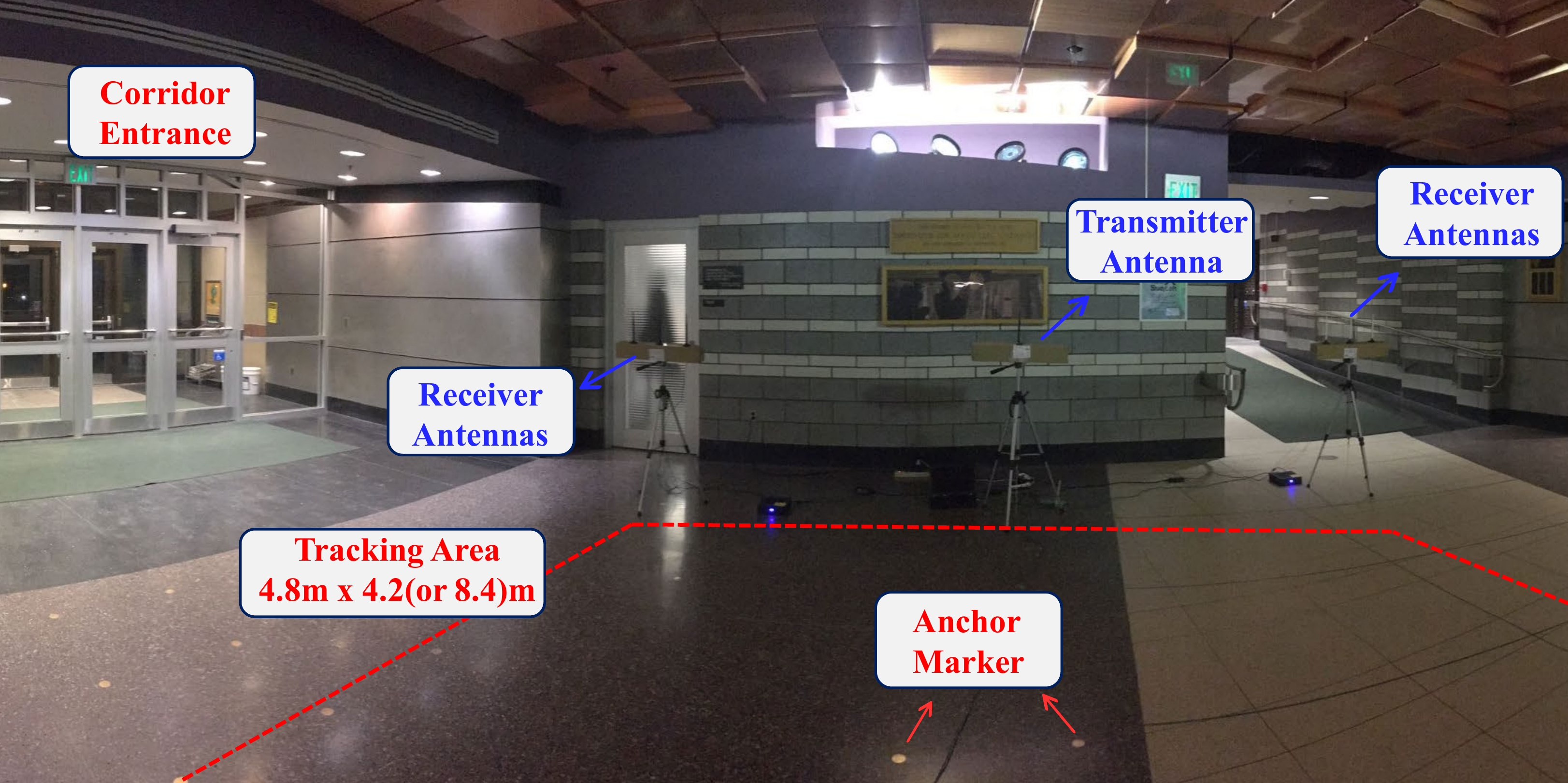}}
\subfigure{\includegraphics[width=0.2\textwidth]{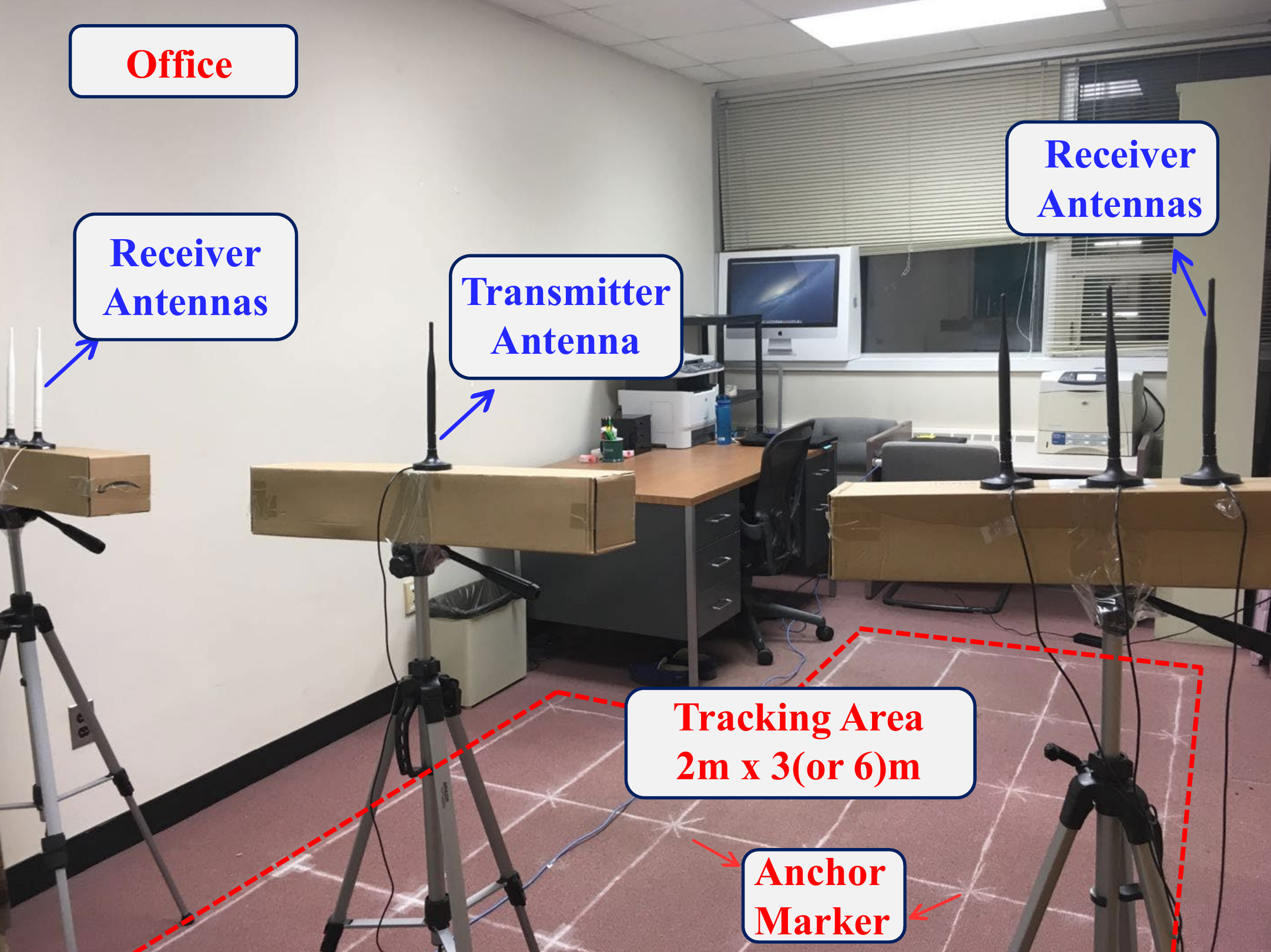}}
\\\vspace{-2mm}
\setcounter{subfigure}{0}
\subfigure[Classroom]{\includegraphics[width=0.23\textwidth]{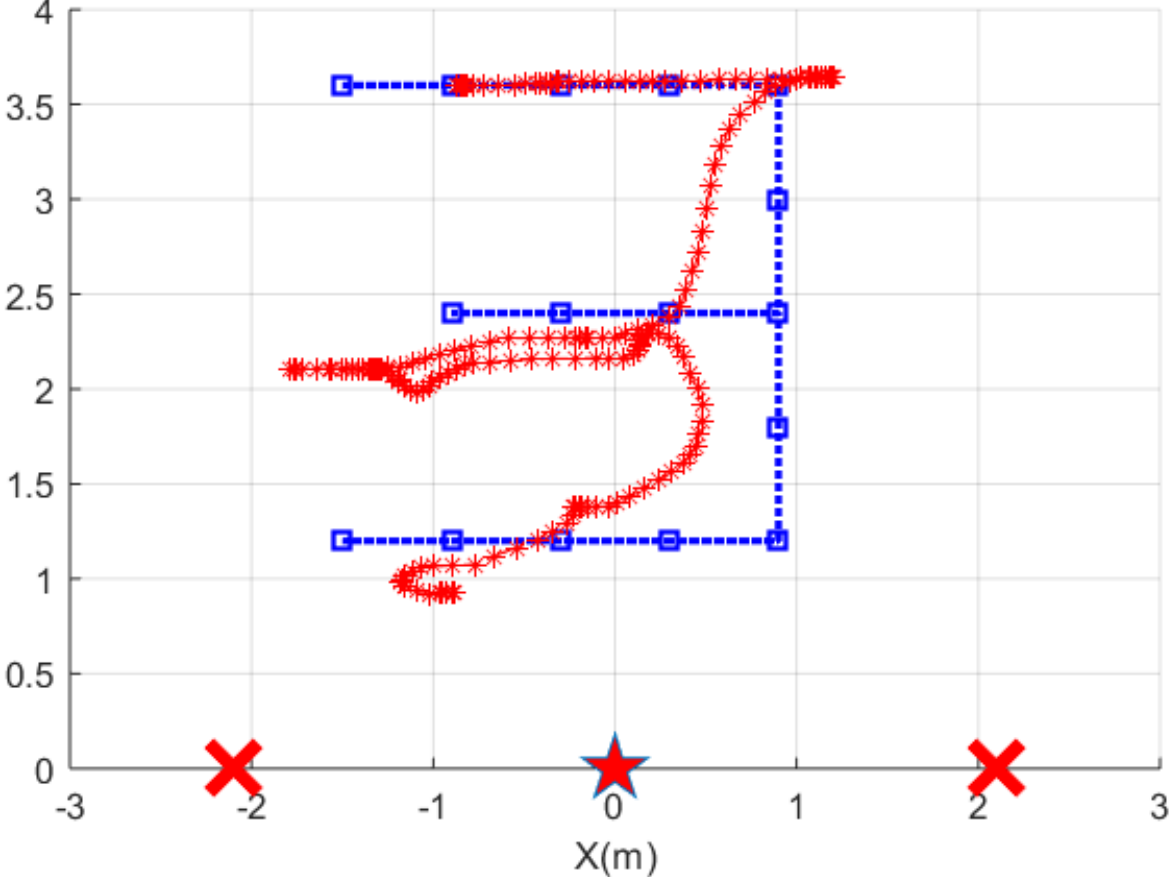} \label{classroom}}
\subfigure[Corridor Entrance]{\includegraphics[width=0.23\textwidth]{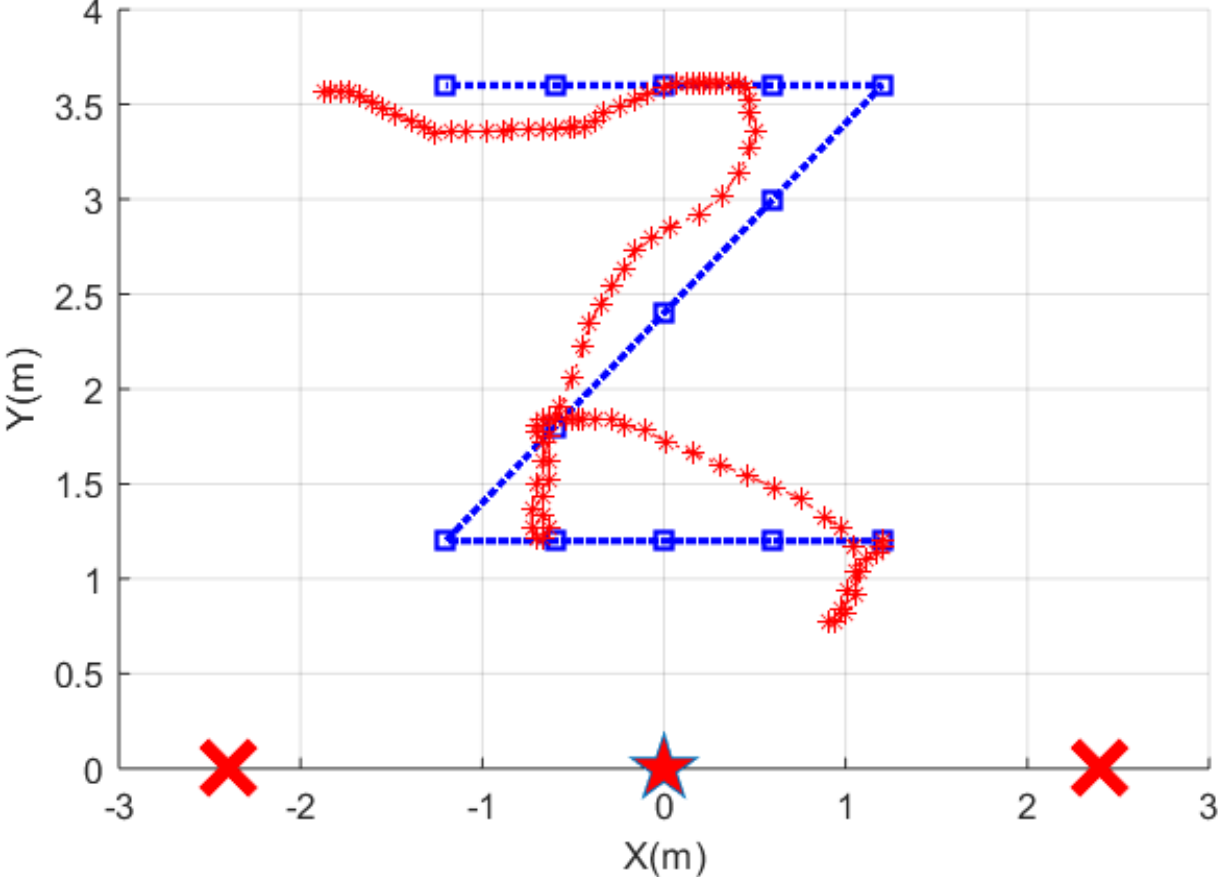}\label{corridor_z}}
\subfigure[Office]{\includegraphics[width=0.23\textwidth]{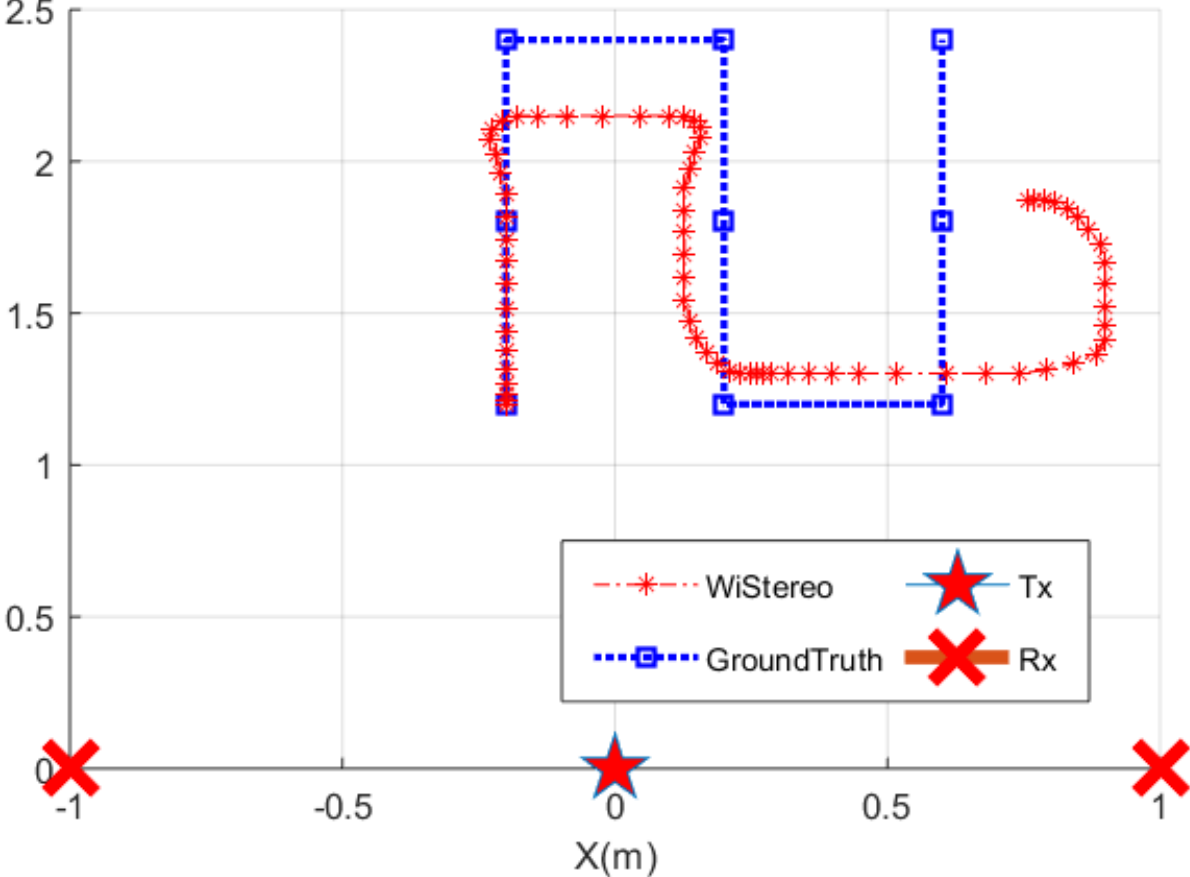} \label{office}}
\vspace{-2mm}
\caption{Experimental setup and corresponding tracking demos in three different environments. The size of the monitoring area in different environments. (a) 3.6m $\times$ 7.2m in classroom, (b) 4.8m $\times$ 8.4m at corridor entrance and (c) 2m $\times$ 6m in office.}
\label{setup}
\vspace{-5mm}
\end{figure*}

Given the walking direction status inferred by the two reference antennas, we determine the walking direction in the mesh model. After we find the landmark position along the walking direction, we need to compute the arrival time at which the user arrives the landmark for walking speed estimation. 
First, we fetch the corresponding auxiliary antennas whose ellipses determine the landmark. For reference antennas $a^1_r$ and $a^2_r$, we use $a^1_a(k)$ and $a^2_a(k)$ to indicate their auxiliary antennas. 
Then, we use the SSD similarity among the reference and auxiliary antennas to calculate the arrival time.

\vspace{1mm}
\noindent
\textbf{Spatial Signal Correlation:} We use the Earth Mover's Distance (EMD)~\cite{EMD} to define the similarity between two SSD vectors $\vec{S}(t,a)$ and $\vec{S}(t',a')$, which is indicated as $EMD((t,a), (t',a'))$. $EMD((t,a), (t',a'))$ illustrates the similarity between these two SSD vectors. The advantage of EMD similarity is that it matches the two distribution represented by all elements of the two SSD vectors. Considering the constant SSD (Equation~\ref{equ-phase-offset}) and random noise in each SSD vector, the distribution similarity is more noise-resilient in comparison with the element-by-element similarity.

\vspace{1mm}
\noindent
\textbf{Speed Calculation:} As shown in Figure~\ref{fig-geometric-model}, we notice that the EMD distance between the SSD vectors observed by reference antennas at time $t_k$ and the SSD vectors observed by auxiliary antennas at time $t_{k+1}$ (arrival time) should be the minimum. Then, the calculation of the arrival time of the landmark can be formulated as a search problem as follows:
\begin{equation}
\label{equ-arrival-time} 
\begin{split}
t_{k+1}=\operatorname*{argmin}_{t_i\in(t_k,t_k+T)}&EMD(\vec{S}(t_k, a^1_r)+\vec{S}(t_k,a^2_r), \\
&\vec{S}(t_i,a^1_a(k))+\vec{S}(t_i,a^2_a(k)))
\end{split}
\end{equation}
where we combine the SSD vectors obtained by two reference antennas to find the most similar combination of the SSD vectors obtained by two auxiliary antennas. We use a fixed time step to search the $t_i$, when we achieve the minimum EMD distance, as the arrival time. After we obtain the arrival time, according to the distance between the start position and the landmark position, we calculate the walking speed. 

\section{Trajectory Recovery}
\label{sec-trajectory-recovery}

We can recursively recover the walking trajectory with continuous walking velocity estimated every $T$ time. 
To further refine the walking velocity error, we notice the walking stability and continuity, which means the walking speed between adjacent estimation periods should not change too much. 
Then, we combine all mesh models and SSD vectors in all walking velocity estimation processes to formulate the global estimation of the arrival time sequence $(t_1, t_2, ..., t_K)$ as an optimization problem as follows:
\begin{align}
    \label{equ-global-optimization}
    &min \sum_{i=1}^K (EMD(t_i)+\omega C(t_{i+1},t_{i})) \\
    &s.t. |t_{i+1}-t_{i}|<T; C(t_{i+1},t_{i})={|t_{i+1}-t_{i}|}/{T} \nonumber
\end{align}

Specifically, for the $k^{th}$ walking velocity estimation, given the search space defined in Equation~\ref{equ-arrival-time} and search time step $\delta_t$, we calculate all the EMD distance for each arrival time candidate to form an EMD distance vector. $EMD(t_i)$ is an element in the EMD distance vector of the $i^{th}$ walking velocity estimation process.
Then, we establish the global EMD distance matrix by joining the EMD distance vectors of $K$ walking velocity estimation processes. 
To derive the arrival time sequence $(t_1, t_2, ..., t_K)$, $C(t_{i+1},t_{i})$ is the cost for arrival time changes from the $i^{th}$ walking velocity estimation process to the $t_{i+1}$. It is defined as ${|t_{i+1}-t_{i}|}/{T}$ and $\omega$ is a positive weighting factor for the cost. The cost function reflects the walking speed changes. To minimize the EMD distance as well as the cost, we punish jumpy peaks of arrival time between adjacent walking velocity estimation processes. 
In this way, we can derive a stable arrival time sequence. We notice that the $t_{i+1}$ is only related to the selection of $t_{i}$, then we resolve the optimization problem with dynamic programming. After we obtain the arrival time sequence, we update the walking speed in each walking velocity estimation process, then refine the walking trajectory.

\section{Implementation and Evaluation}
\label{sec-implementation-and-evaluation}

\begin{figure*}[!t]
    \centering
    \subfigure[Localization Error]{\includegraphics[width=0.27\textwidth]{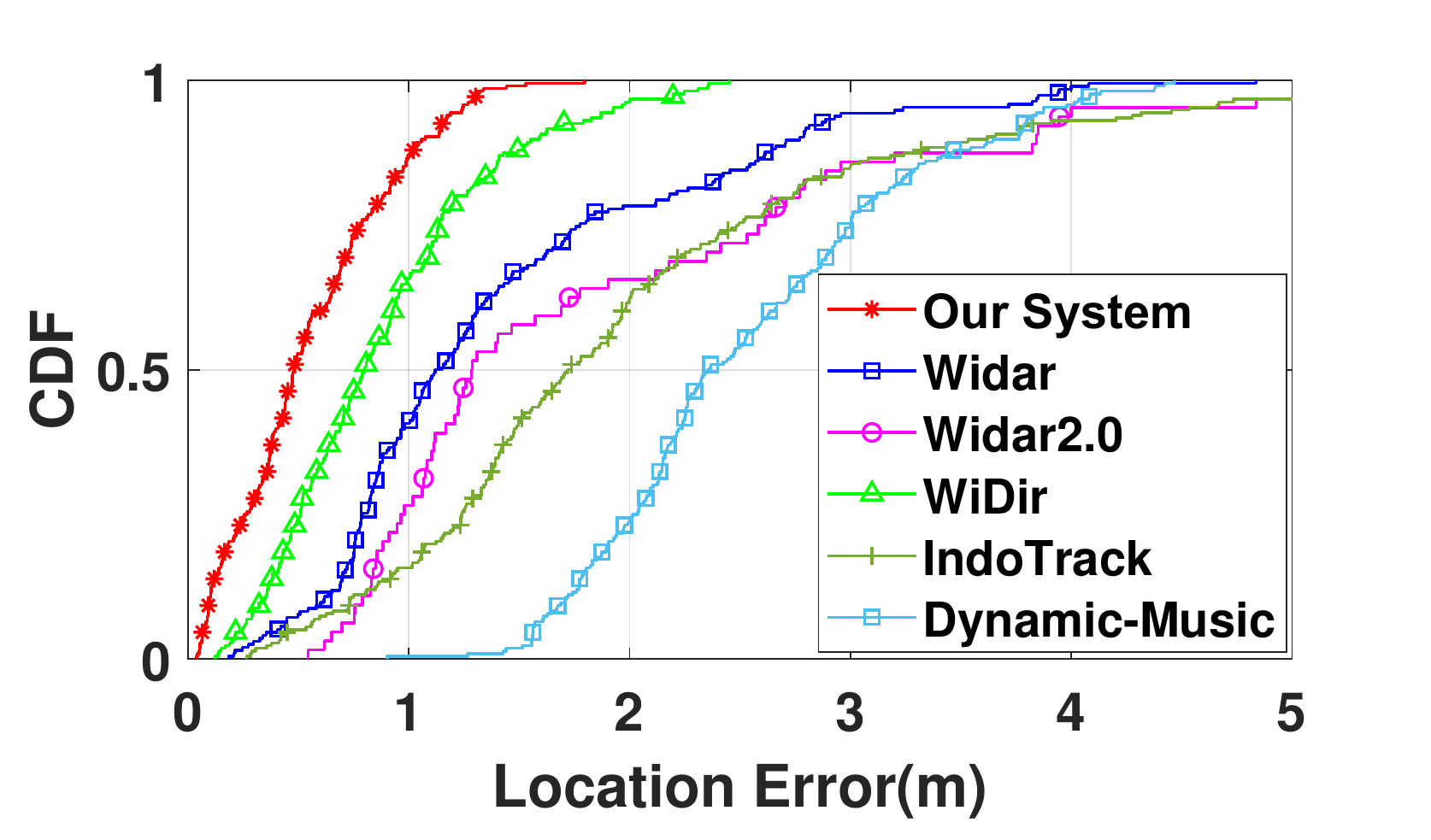}\label{comparison}}
    \hspace{-.1in}
    \subfigure[Different Environment]{\includegraphics[width=0.27\textwidth]{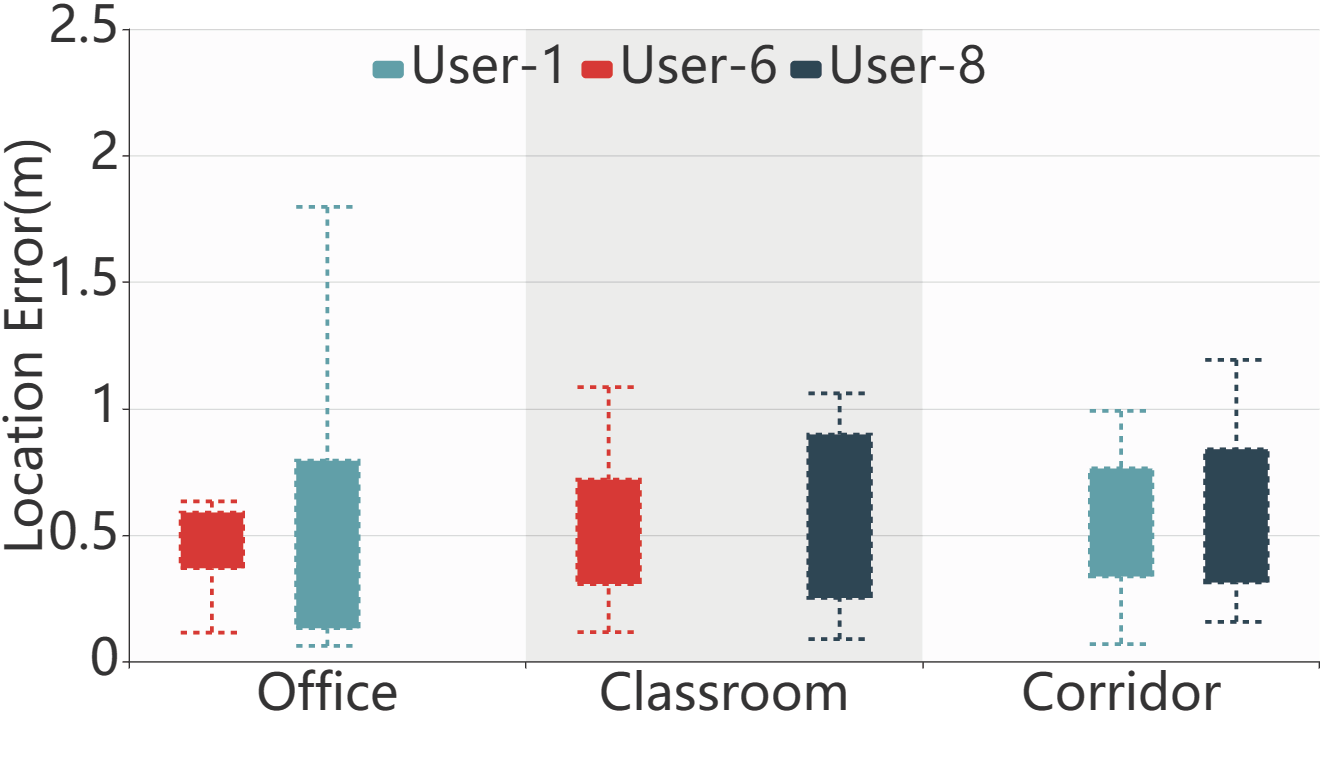}\label{robust}}
    \hspace{-.1in}
    \subfigure[Different Modules]{\includegraphics[width=0.27\textwidth]{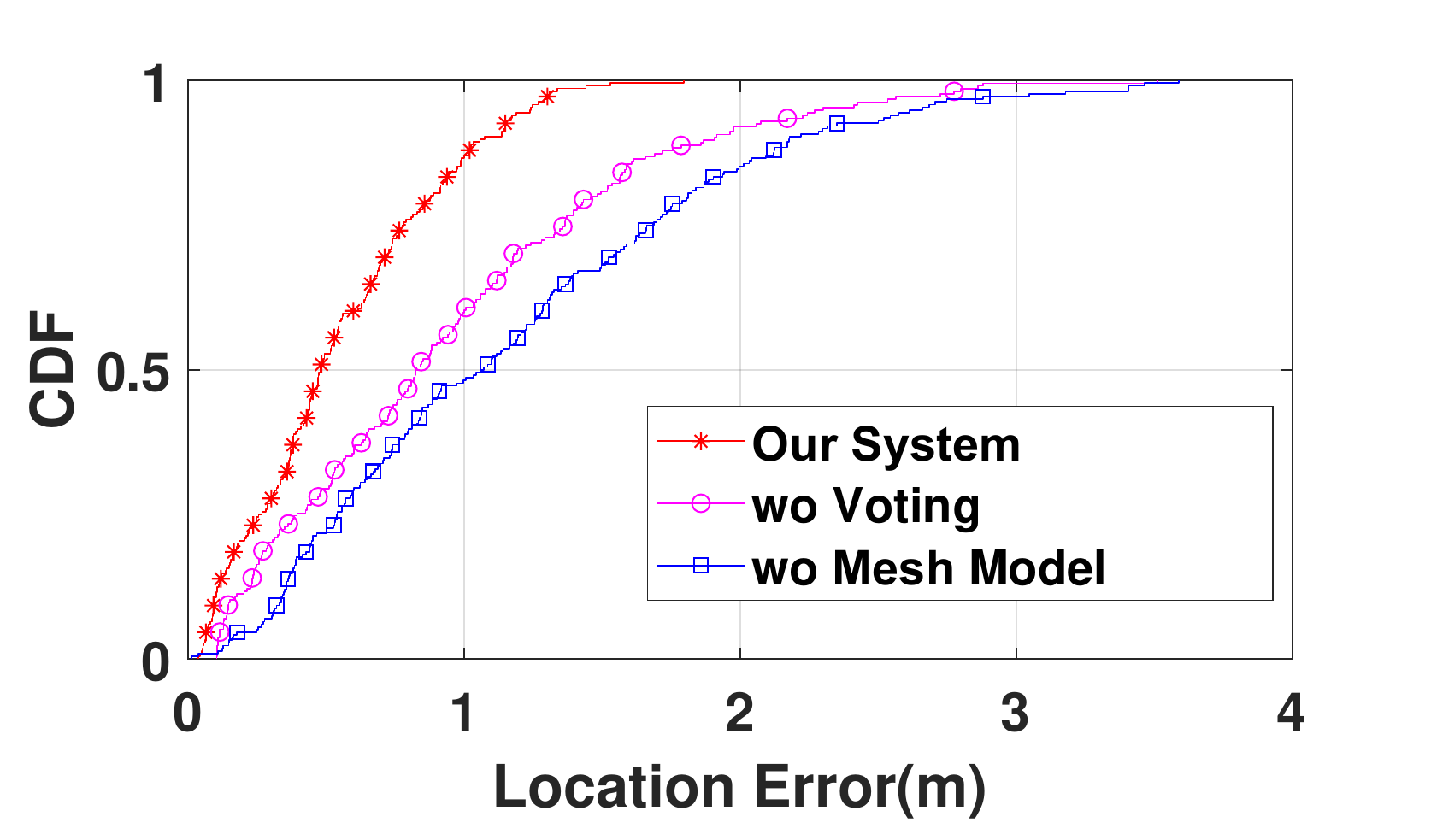}\label{module}}
    \vspace{-2mm}
    \caption{Overall Performance of \ours, including (a) comparative study with state-of-the-art approaches, (b) performance robustness across users and scenarios, (c) effectiveness of different functional modules of \ours.}
    \label{overall_performance}
    \vspace{-5mm}
\end{figure*}

\subsection{Implementation}

We have implemented \ours using only one transmitter with one antenna and two receivers with three antennas, including off-the-shelf mini-desktops and laptops equipped with an Intel 5300 NIC. Linux CSI Tool~\cite{csitools} is installed to acquire CSI measurements. Two receivers are set to monitor the packets broadcast by the only antenna on the transmitter at a rate of 1,000 packets per second. To alleviate the radio interference, we select channel 165 at 5.825 Ghz~\cite{channel}. We place the two transceiver pairs as the ($\S$\ref{sec-geometrical-model}) mentioned to form the mesh model.


\noindent
\textbf{Experimental Setup.} We conduct experiments in 3 indoor scenarios to fully evaluate the performance of \ours: a large venue at the corridor entrance, a small office with kinds of devices (e.g., desktop, printer), and a crowded classroom full of desks and chairs. The sizes of the monitoring areas are 3.6m$\times$7.2m, 4.8m$\times$8.4m, and 2m$\times$6m in the classroom, corridor entrance, and office, covering the symmetric areas at the two sides of the antenna plane. Figure~\ref{setup} demonstrates the deployment of transceivers and corresponding coverage, respectively. All seven antennas are aligned and carefully deployed to generate the mesh model in the monitoring area.

\noindent
\textbf{System Parameters.} The searching time period $T$ is 128ms. The distance between the transmitter antenna and the two reference receiver antennas is 2.1m, 2.4m, and 1m in the classroom, corridor entrance, and office. The distance between the two receiver antennas is 20cm. The threshold of sign distance used to judge whether the target stays at the current ellipse is 4.5. The length of the time window and packet delays are 32ms and 5ms used to construct the virtual voters. The number of time windows is 4 in a time period.

\noindent
\textbf{Dataset.} We collect multiple trajectories from 3 scenarios concerning with 6 different volunteers and 18 types of trajectories. 
All experiments are followed by an IRB exception. 
Specifically, the target starts to move about 1s behind the start of data recording and we terminate the recording after the target arrives at the destination about 1s. All 18 trajectories consist of common ones that occurred in real life, including straight line, "L" line, "U" line, "S" line, "M" line, "Z" line, and arc. And the length of the trajectories spans from 2.83m to 10.8m. This dataset contains 216 trajectory instances (3 scenarios$\times$3 users$\times$6 trajectories$\times$4 instances). 
%

\noindent
\textbf{Ground Truth.} To obtain real trajectory coordinates as the ground truth, some noticeable markers are attached and connected to form a real meshed space as denoted in Figure~\ref{setup}. First, volunteers are asked to stand at the assigned location and keep aligning the vertical middle line of the torso with the gridline while walking. No extra requirements are needed on the speed or gait for volunteers except stopping at the pre-defined destination denoted with a marker. To align the timestamps of the ground truth and the derived trajectory from \ours while alleviating the fluctuation of proceeding velocity, we employ interpolation and down-sampling to both trajectories. Then Dynamic Time Wrapping (DTW) distance is calculated and averaged for each sampling point to measure the tracking accuracy. The sampling rate and the DTW window are both set 50 empirically for the location accuracy metric.

\subsection{Trajectory Recovery Overview}
\label{subsec-trajectory-recovery-overview}

Instead of indirectly measuring the simultaneous walking speed, we evaluate the trajectory recovery. To demonstrate the properties of \ours visually, we first show some trajectory recovery results from various types of trails in different scenarios. Illustrated in Figure~\ref{setup}, the straightforward cognitive has three folds. 
\begin{enumerate}[wide, labelwidth=!, labelindent=0pt]
    \item \ours is sensitive enough to detect the rotation, even the rotating angle is large (e.g. $180^\circ$ in the middle left marker of "M" line in the classroom, $135^\circ$ in the two turning points of "Z" line in the corridor, and $90^\circ$ in the up-left marker of "S" line in office)~(See $\S$\ref{subsec-parameter-study}). It attributes to the two-step SSD sign based direction estimation process, rendering a more reliable turning points detection. Massive virtual voters also play a key role to determine the direction collaboratively.
    \item Noticeable deviation occurs when the target moves to the edge of the monitoring area (especially areas on both sides), which stays away from the link, like the up-left marker of the "Z" line in the corridor and down-right marker of "S" line in the office. The reason is that the distribution of the new position candidates in the mesh model is not uniform in all monitoring areas. At those edge spaces, the mesh model generated by crossing ellipses can transform severely and the spacing between two adjacent new position candidates gets smaller, rendering a larger walking speed estimation error.
    \item It can be derived that the trajectory is more robust when people move in the central part and it has more reliable estimations instead of staying far away from the link, as illustrated in Figure~\ref{corridor_z}. The reason is the new positions are much more uniform in the central area due to the symmetric distribution compared with the ones in edge areas~(See $\S$\ref{subsec-parameter-study}). 
\end{enumerate}

\begin{figure}[!t]
    \centering
    \subfigure[Tracking Duration]{\includegraphics[width=0.23\textwidth]{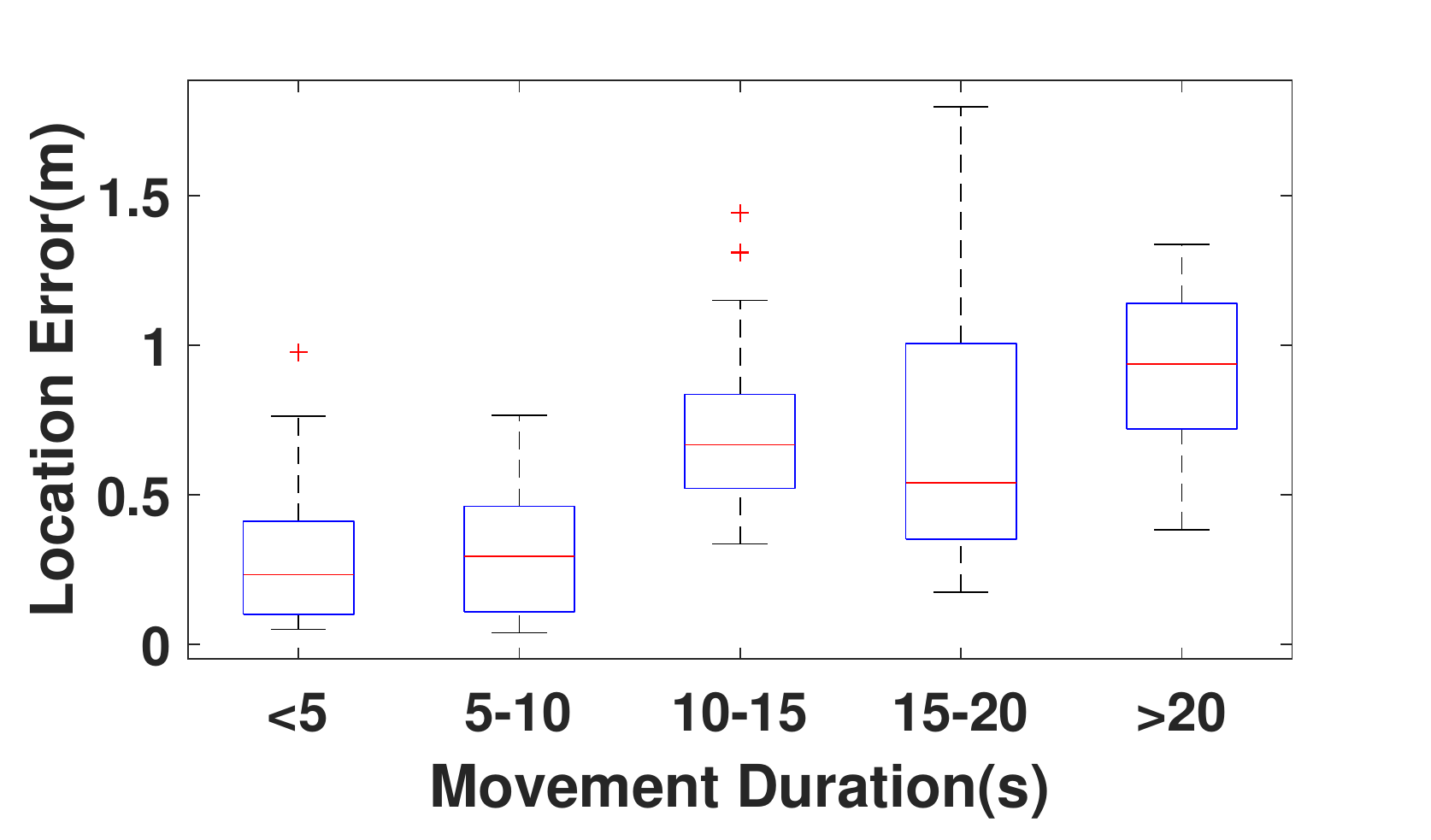}\label{duration}}
    \subfigure[Different Persons]{\includegraphics[width=0.23\textwidth]{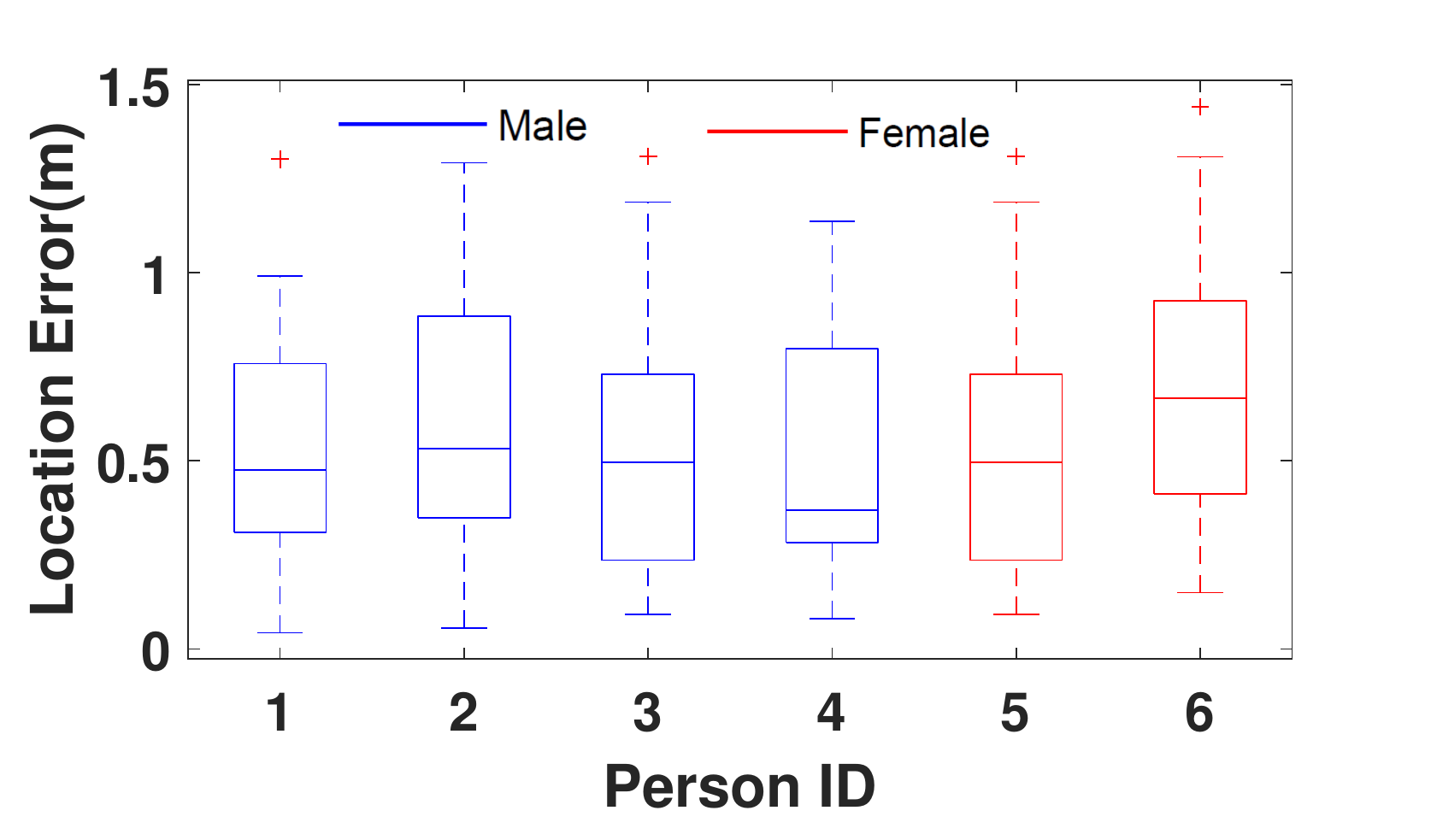}\label{person}}
    \subfigure[Different Scenarios]{\includegraphics[width=0.23\textwidth]{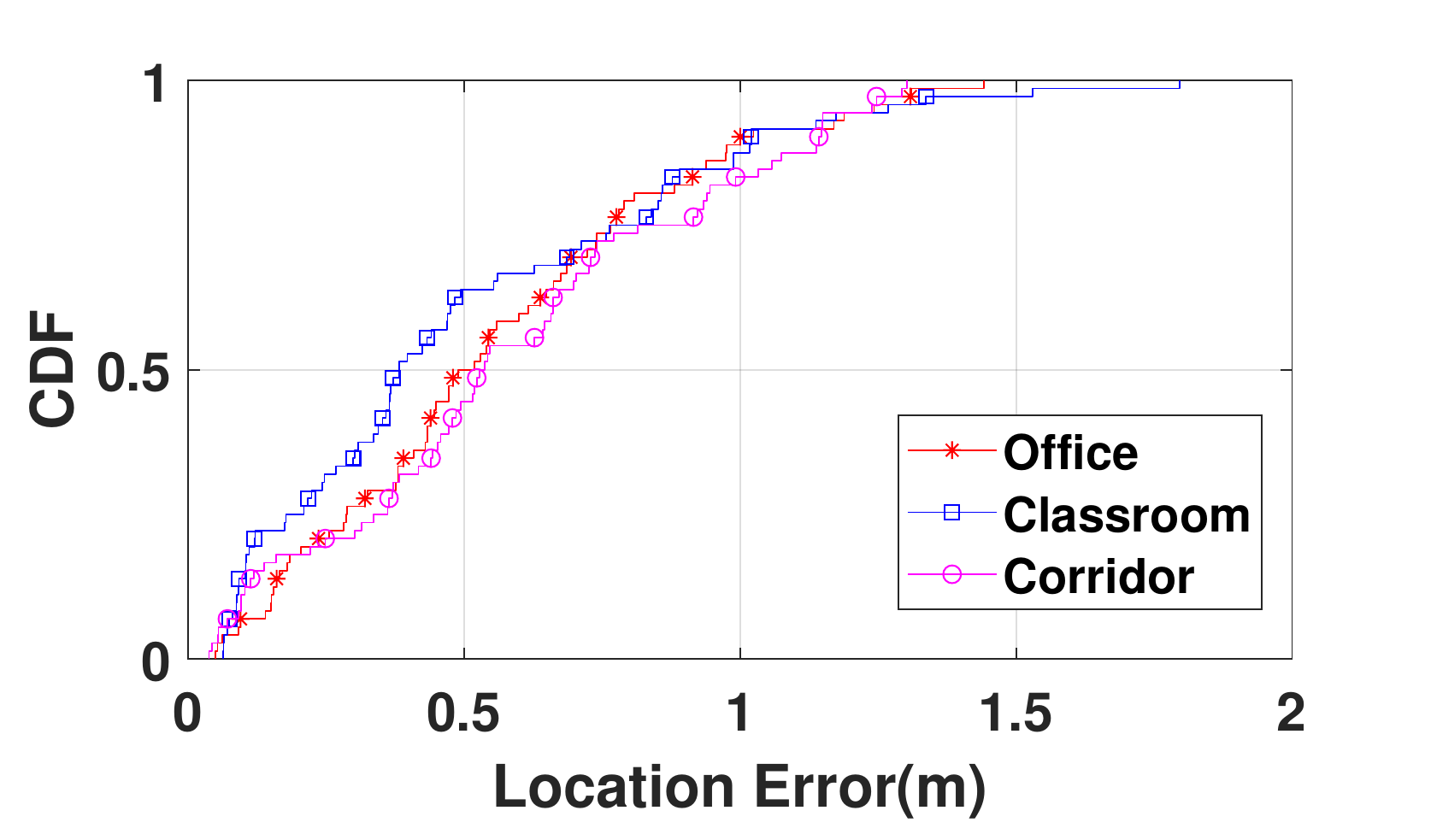}\label{scenario}}
    \subfigure[Mesh Model Distribution]{\includegraphics[width=0.23\textwidth]{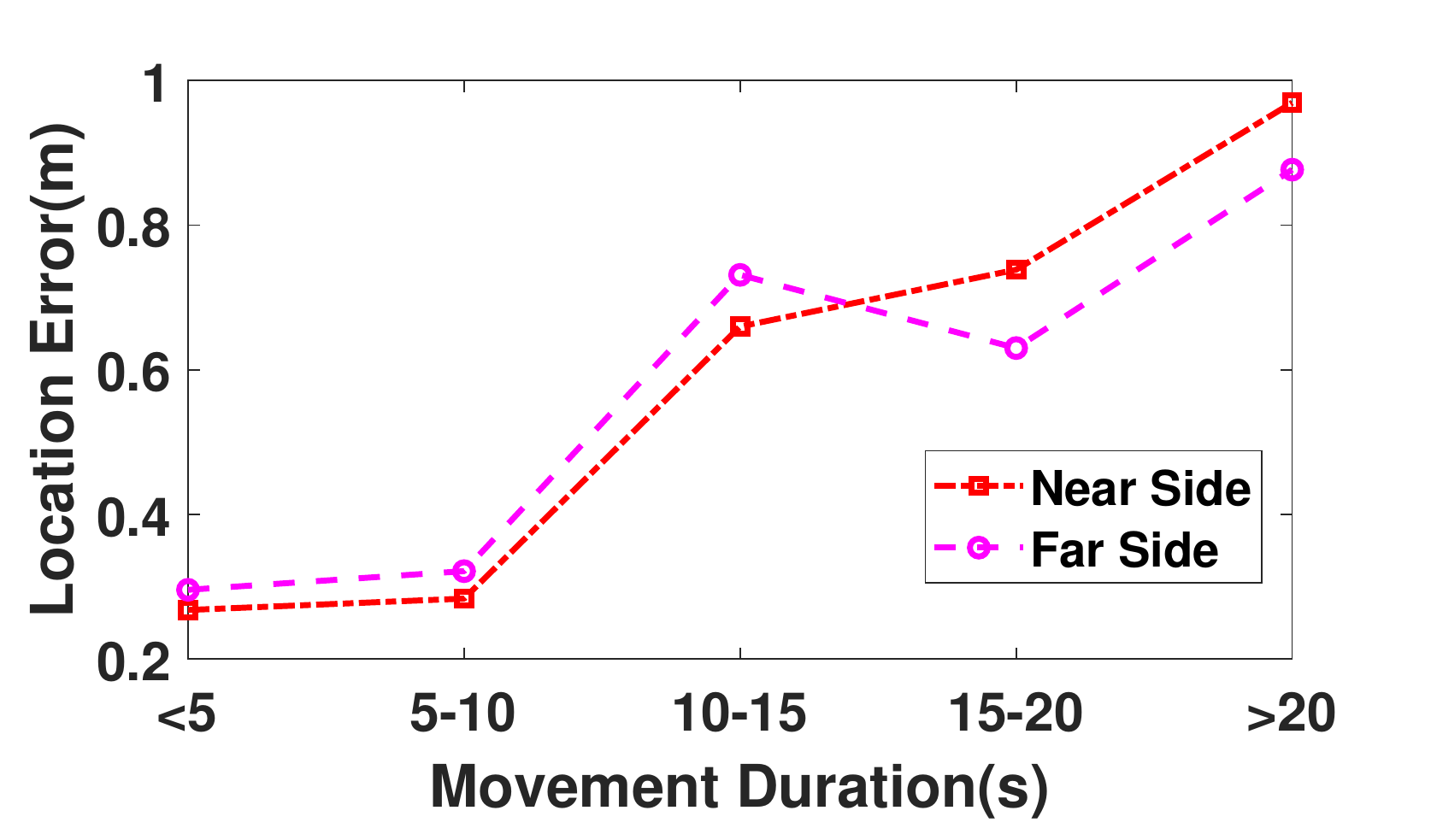}\label{near_far}}
    \subfigure[Different Tracking Areas]{\includegraphics[width=0.23\textwidth]{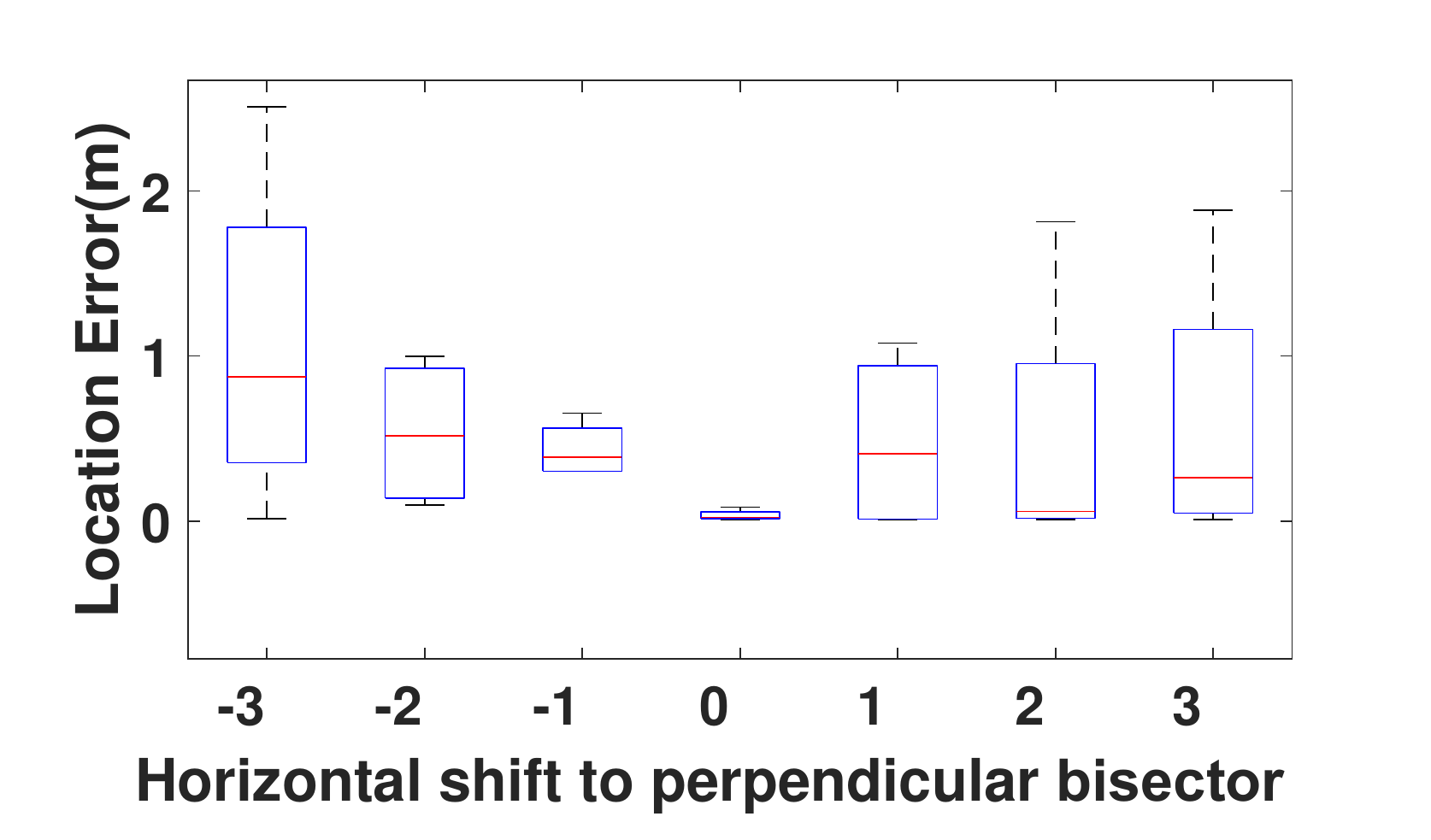}\label{straight_trails}}
    \subfigure[Different Trajectories]{\includegraphics[width=0.23\textwidth]{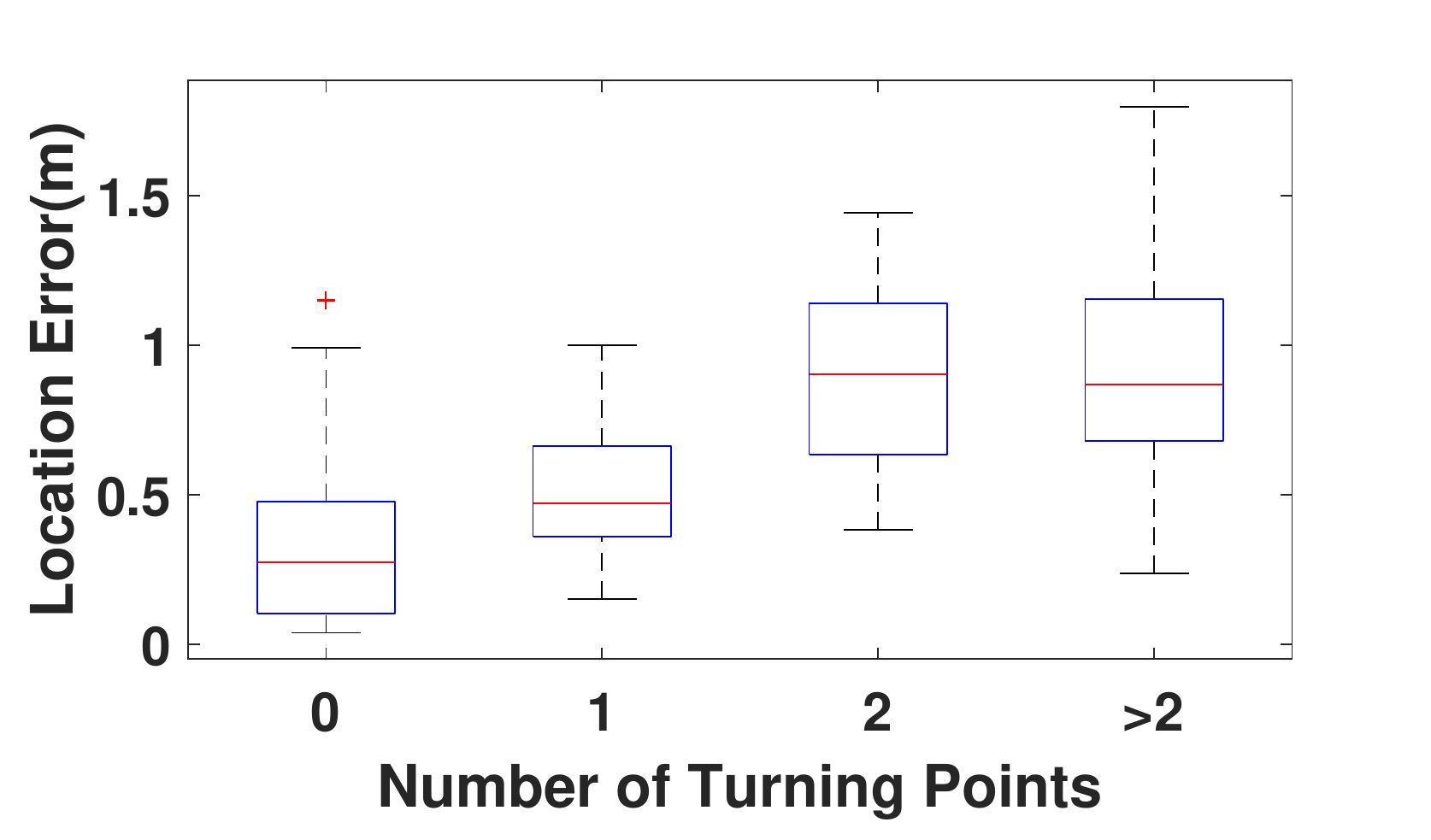}\label{angel_resolution}}
    \vspace{-2mm}
    \caption{The impact of the diverse (a) tracking duration, (b) persons, (c) scenarios, (d) mesh model distribution, (e) tracking areas with different horizontal shifts to perpendicular bisector, (f) trajectories with the different number of turning points.}
    \label{parameter_study}
    \vspace{-7mm}
\end{figure}

\subsection{Overall System Performance}
\label{subsec-overall-system-performance}

Next, we show the overall system performance of \ours covering all 6 participated volunteers in three scenarios. As shown in Figure~\ref{comparison}, \ours achieves a median localization DTW error of 0.47m and 90-percentile error is 1.06m.

\vspace{1mm}
\noindent
\textbf{Comparative Study.} We first compare \ours with five state-of-the-art works (e.g., Widar \cite{Widar}, Widar2.0 \cite{Widar2.0}, WiDir \cite{WiDir}, IndoTrack \cite{IndoTrack}, Dynamic-Music \cite{Music}).
Specifically, Widar \cite{Widar} and WiDir \cite{WiDir} rely on the mesh model for passive tracking, in which the former associates the Wi-Fi measurements with the moving velocity and locations while the latter only predicts the coarse-grained walking direction with the majority sign of an SSD vector. And we supplement our mesh model to WiDir, enabling it for passive tracking. Besides, the other three works~\cite{Music,IndoTrack, Widar2.0} employ the range parameter estimation. For example,  Dynamic-Music \cite{Music} leverages JADE to estimate AoAs of signals bouncing off the target and pinpoints the intersection of AoAs for localization. IndoTrack \cite{IndoTrack} further incorporates DFS with AoA by calculating the target's velocity from instantaneous DFS. And Widar2.0~\cite{Widar2.0} only requires a single link by jointly estimating all range parameters.

For fairness, we utilize the two transceiver pairs for all approaches. The results are shown in Figure~\ref{comparison}. 

First of all, we compare the tracking performance between \ours and existing passive tracking methods. The 90-percentile error of \ours is 1.06m while it can reach 4m for Widar~\cite{Widar}, Dynamic-Music~\cite{Music}, IndoTrack~\cite{IndoTrack}, and Widar2.0~\cite{Widar2.0}, which is approximately 3 times larger than \ours. In our experiments, we have some complex paths (e.g. M line, S line), which contain multiple turns. The existing passive tracking methods are not sensitive enough to accurately detect those turns so they incur relatively large accumulated errors in the worst cases. In contrast, the location error of \ours is well controlled with our fine-grained mesh model based walking velocity estimation. Moreover, the median error of \ours is 0.47m which also outperforms the existing passive tracking methods. Among the existing passive tracking methods, the median errors of Widar and Widar 2.0 are 1.15m and 1.29m which are better than that of Dynamic-Music~\cite{Music} and IndoTrack~\cite{IndoTrack}. 

Second, we compare the tracking accuracy between \ours and WiDir. We can see the location error of WiDir increases more quickly than \ours. The 90-percentile and median errors of WiDir are 1.59m and 0.8m. The main reason is the walking direction estimation of WiDir relies on the majority sign of SSD vector which is more vulnerable facing the residual SSD noise. In contrast, both sign distance and massive voting are noise-resilient so they provide more accurate walking direction estimation. Moreover, the tracking accuracy of WiDir is better than the other range parameter based methods. The reason is that the implementation of WiDir adopts the mesh model and walking speed calculation of \ours, which can bound the accumulated tracking errors even experiencing certain walking direction errors.

\ours outperforms Widar and Widar2.0, which have $1.26m$ and $2.27m$ for the DTW location error, respectively. It's mainly because that Widar leverages the time-frequency analysis for PLCR extraction, which suffers from the multi-path effects. Besides, direction determination is a serious problem for Widar, especially for a complex and lasting movement such as the "S" line or "M" line. Because it only fuses the opportunistic calibration with WiDir~\cite{WiDir} and fails to recognize the direction in a robust and fine-grained scale. That's why the author notes Widar needs at least three receivers for accurate direction estimation. However, \ours achieves reliable direction estimation by designing the SSD feature and a two-step scheme for direction estimation. Moreover, calling on massive virtual voters makes the estimation more robust. Given Widar2.0 only utilizes one link, it collects less information even if using jointly multi-dimensional parameter estimation, resulting in a larger location error for human tracking. Widar2.0 further demonstrates the sensitivity and instability of the range-based methods to the dynamic environments

\vspace{1mm}
\noindent
\textbf{Performance across Users and Scenarios.} Then we evaluate the robustness of \ours by analyzing various users and scenarios. Figure~\ref{robust} shows median localization errors for all three users are almost identified across two different scenarios, less than 0.5m. Given the most variance of 0.2m for user-8 between the classroom and corridor, it is acceptable with the size of the human body ($0.3\sim 0.5 m$ in width). Hence, we conclude that \ours demonstrates relatively consistent performance with different users across scenarios.

\vspace{1mm}
\noindent
\textbf{Benefits of Individual Modules} We evaluate the effectiveness of the proposed mesh model ($\S$\ref{sec-geometrical-model}) and massive voting strategy ($\S$\ref{sec-walk-direction}), shown in Figure~\ref{module}. First, we remove the massive virtual voting process, rending a coarse-grained direction estimation. The performance suffers and leads to a doubled median location error of 0.85m. Without the mesh model, the median location error gets much worse, reaching more than 1m. Meanwhile, the 90-percentile location errors increase more quickly without either voting or mesh model. Since it only has direction information and cannot determine the target's location without reliable distance measurements. The intuition is that fine-grained direction estimation enables an effective walking direction estimation, and the mesh model derives an accurate moving distance along this reliable direction, both are indispensable.

\subsection{Detailed Parameter Study}
\label{subsec-parameter-study}

We further evaluate the influence of different user experiences and parameters on the performance of \ours.

\vspace{1mm}
\noindent
\textbf{Impact of Movement Duration.} Error accumulation is a serious problem for human tracking. To evaluate the impact of the moving duration, we first divide the trails into five groups for different movement duration, namely $<$5s, [5,10)s, [10,15)s, [15,20)s, $\geq$20s. Then we calculate the corresponding DTW error distribution. As Figure~\ref{duration} shows, tracking error statistically increases as targets proceeds, but the worst case is still controllable. When the target moves more than 20s, the most of tracking error is less than 1.5m. However, we can find the tendency is not linear rigidly. For example, $<5s$ and $[5,10)s$ have a similar location error while $[15,20)s]$ has a smaller median error compared with the adjacent groups. And it demonstrates \ours can alleviate the error accumulation. To further solve the problem, we can leverage opportunistic calibration~\cite{Widar}, which utilizes an explicit indoor map to calibrate the trajectory, for the tracking process.

\vspace{1mm}
\noindent
\textbf{Impact of Human Diversity.} To determine the effect of target diversity, we recruit 6 volunteers with different genders, body shapes, and heights, to evaluate their influence on the performance of \ours. Each time volunteers are explained with basic experimental settings and tips before the experiment, such as the marker deployment and trail routes as the ground truth. And they are not specially required and only expected to walk in normal conditions. Figure~\ref{person} plots the localization errors of \ours with different people. And we can see the median localization error of all users is similar around 0.5m, no more than 0.7m for all users. Hence, \ours keeps resilient to human diversity and achieves consistent tracking accuracy across all volunteers.

\vspace{1mm}
\noindent
\textbf{Impact of Scenario Diversity.} To demonstrate the influence of scenario diversity, three different scenarios are tested comprehensively including office, classroom, and corridor entrance. As shown in Figure~\ref{scenario}, \ours achieves low median localization errors of 0.49m, 0.39m, and 0.52m, respectively. The performance is much similar. This verifies that \ours is resilient to multi-path effects or fluctuated noises, leading to ubiquitous application in different indoor environments. However, the worst-case errors in the classroom and office are larger than those in the corridor even it has a larger tracking area. The reason is that the furniture in the classroom and office leads to more signal noises.

\vspace{1mm}
\noindent
\textbf{Impact of Mesh Model Uniformity.}
Given the mesh model, the distance among new positions usually changes in different monitoring areas, rendering a non-uniform distribution of the new positions for \ours. Thus we do experiments to explore different tracking areas, vertically and horizontally.
To evaluate the performance of various horizontal tracking areas, near or far from transceivers, we divide the trails into two groups, one has the starting point near transceivers while the other has the one at the far side. Since the larger reflecting body part at the near-side can induce multiple reflected paths, it may lead to erroneous motion determination at the beginning. Illustrated in Figure~\ref{near_far}, we find \ours shows consistent location error with the near and far stating points for various movement duration. It demonstrates \ours is resilient to the impact of different starting points on the error accumulation and has a comparative performance with horizontal tracking areas near or far from transceivers.

We further explore the vertical tracking areas, near or far from the perpendicular bisector of transceivers. And seven vertical trails to the transceivers are distributed with equal spacing horizontally. Figure~\ref{straight_trails} shows that the perpendicular bisector trail has the best performance, reaching around $5cm$ location error. And the error increases with the distance to the perpendicular bisector. The reason is the trails near the perpendicular bisector of the transceivers as the symmetrical distribution of the new positions, leading to more reliable direction estimation. Conversely, the trails away from the perpendicular bisector of the transceivers locate at the edge of the monitored area and have non-uniform distributed new positions in the mesh model. We can improve the reliability of \ours using more than two pairs of transceivers, rendering more uniform and fine-grained new positions.

\vspace{1mm}
\noindent
\textbf{Impact of Turning Points.}
To evaluate the angle resolution of \ours for direction sensitivity, we evaluate the performance of trails with various turning points. Illustrated in Figure~\ref{angel_resolution}, tracking error increases with more turning points. However, we find the median location error decreases for the $>2$ group, which has much more complex trails, such as $M$ and $S$ with five or four turning points. Thus the increased location error is mainly caused by the accumulation error with more movement duration. And \ours can be resilient and sensitive to the direction variance (See ($\S$\ref{subsec-trajectory-recovery-overview})).

\section{Related Work}
\label{sec-related-work}


Our work is broadly related to wireless sensing research, which leverages Wi-Fi signals to infer physical measurements (e.g., direction, velocity)~\cite{WiDir,Widar} or semantic context representatives (e.g., activity, gesture) indoors~\cite{li2020wihf,WiSIA_Sensys20}.

\vspace{1mm}
\noindent
\textbf{Device-based Localization.} Numerous efforts have been devoted to device-based localization in the past decade. Many works resort to parameter estimation (e.g., ToF~\cite{tof-1,aoa+tof}, AoA~\cite{aoa+tof,SpotFi}), leading to sub-centimeter level accuracy with phased arrays and massive Wi-Fi Access Points (APs). Recently, RIM~\cite{RIM} proposes to capture the temporal correlation pattern across spatial antennas and achieve spatial-temporal virtual antenna retracing, leading to noise-resilient human tracking. However, all of these methods require users to carry devices as transceivers, making them not feasible or even impossible for some scenarios. Due to the limitation of intrusive deployment and privacy issues, \ours aims to achieve accurate human tracking in device-free scenarios.

\vspace{1mm}
\noindent
\textbf{Device-free Tracking.} To suppress the interference on the weak human reflections in device-free scenarios, many methods are proposed for tracking more accurately and reliably, including dedicated hardware~\cite{WiTrack,WiTrack2.0}, specially designed statistical model~\cite{LiFS,karanam2019tracking,Widar}, or jointly multi-parameter estimation~\cite{IndoTrack,Widar2.0,md-Track}. Specifically, LiFS~\cite{LiFS} models the fine-grained subcarrier information of CSI measurements with the power fading model, while Widar~\cite{Widar} proposes the CSI-mobility model associating Path Length Change Rate estimated by Doppler Frequency Shift (DFS)~\cite{CARM} with walking velocity. Besides, Karanam et al.~\cite{karanam2019tracking} refine estimated motion parameters with multiple filters using the magnitude of Wi-Fi measurements, enabling tracking 1-3 people. All statistical models require multiple Wi-Fi NICs to collect Wi-Fi measurements from various spatial vantage views. To reduce the deployment cost, the jointly multi-parameter estimation has been adopted, especially Widar2.0~\cite{Widar2.0} and md-Track~\cite{md-Track} only demand a single transceiver pair by employing all parameters such as ToF, AoA, DFS, and PA. However, it suffers from the accumulated tracking errors, especially for complex paths.


\section{Conclusion}
\label{sec-conclusion}

In this paper, we propose \ours which enables fine-grained walking velocity estimation so that achieves accurate passive human tracking with COTS Wi-Fi hardware. In comparison with existing approaches which rely on DFS, \ours uses two Wi-Fi transceiver pairs to design a mesh model, then utilize temporal-spatial signal correlation features to achieve error-bounded walking direction and speed estimation.
Additionally, we design several methods to avoid the direction and speed errors incurred by multi-path signal noises. 
We implement \ours prototype with three laptops equipped with Intel 5300 NIC and evaluate its performance through intensive experiments in different environments with different people. The experimental results show that the median error of human tracking is 0.47m and 90-percentile error is 1.06m, which outperforms state-of-the-art systems. In the future, we will test \ours in larger areas and develop an enhanced system using temporal-spatial signal correlation to estimate the walking velocities of multiple persons simultaneously. 

\section*{Acknowledgment}
We would like to thank our anonymous reviewers
for providing valuable feedback on our work. This work was supported in part by National Science Foundation grants CNS-1909177.

\bibliographystyle{IEEEtran}
\bibliography{reference} 

\end{document}